\documentclass{amsart}

\usepackage{graphicx}
\newtheorem{theorem}{Theorem}

\newtheorem{proposition}{Proposition}[section]

\theoremstyle{definition}

\newtheorem{example}{Example}

\theoremstyle{remark}
\newtheorem{remark}{Remark}[section]

\numberwithin{equation}{section}

\def\d{\text{d}}
\def\e{\text{e}}
\def\i{\text{i}}
\newcommand{\supp}{\operatorname{supp}}
\newcommand{\Real}{\operatorname{Re}}

\begin{document}

\title[products of random matrices]
{Explicit invariant measures for products of random matrices}

\author{Jens Marklof}
\address{School of Mathematics\\
        University of Bristol\\
        Bristol BS8 1TW, United Kingdom}
\email{j.marklof@bristol.ac.uk}

\author{Yves Tourigny}
\address{School of Mathematics\\
        University of Bristol\\
        Bristol BS8 1TW, United Kingdom}
\email{y.tourigny@bristol.ac.uk}

\author{Lech Wo{\l}owski}
\address{School of Mathematics\\
        University of Bristol\\
        Bristol BS8 1TW, United Kingdom}
\email{l.wolowski@bristol.ac.uk}

\thanks{The authors gratefully acknowledge the support
of the Engineering and Physical Sciences Research Council
(United Kingdom) under Grant GR/S87461/01 and an Advanced Research
Fellowship (JM)}





\begin{abstract}
We construct explicit invariant measures for a family of infinite products of random, independent, 
identically-distributed elements of $\text{SL}(2,{\mathbb C})$. The 
matrices in the product are such that one entry is gamma-distributed 
along a ray in the complex plane.
When the ray is the positive real axis, the products are those associated with a continued
fraction studied by Letac \& Seshadri
[{\em Z. Wahr. Verw. Geb.}
{\bf 62} (1983) 485-489], who showed that the
distribution of the continued fraction
is a generalised inverse Gaussian.
We extend this result by finding the 
distribution for an arbitrary ray in the complex right-half plane, and thus 
compute the corresponding Lyapunov exponent explicitly.  
When the ray lies on the imaginary axis, the matrices
in the infinite product coincide 
with the transfer matrices associated with a one-dimensional discrete Schr\"{o}dinger
operator with a random, gamma-distributed potential. Hence, the explicit knowledge of the
Lyapunov exponent may be used to estimate the (exponential) rate of localisation
of the eigenstates.
\end{abstract} 

\maketitle

\section{introduction}
\label{introduction}

Let $\{ {\mathcal A}_{n}\}_{n \in {\mathbb N}}$ denote a sequence of independent random matrices
identically distributed according to a probability measure $\mu$ on $\text{GL}(2,{\mathbb R})$.
The problem of determining the asymptotic behaviour of the product
\begin{equation}
{\mathcal M}_n := {\mathcal A}_1  {\mathcal A}_2 \cdots {\mathcal A}_n \quad \text{as $n \rightarrow \infty$},
\label{productOfRandomMatrices}
\end{equation}
plays a crucial r\^ole in the theory of products of random matrices and its applications, 
especially in mathematical physics \cite{BoLa,CaLa}. 

The rate of growth of this product can be quantified by  
its {\em Lyapunov exponent}
\begin{equation}
\lambda := \lim_{n \rightarrow \infty} \frac{1}{n} {\mathbb E} \left ( \ln | {\mathcal M}_n | \right ),
\label{lyapunovExponentDefn}
\end{equation}
where $| \cdot |$ denotes some matrix norm.

The Lyapunov exponent exists whenever ${\mathbb E}(\log^+|A_{1}|)<\infty$. The Furstenberg--Kesten 
theorem (\cite{FK}, \cite{BoLa} p. 11) generalises the classical strong law of large numbers to the case of 
non-commuting random products and states that
$$
\lambda = \lim_{n \rightarrow \infty} \frac{1}{n} \ln  | {\mathcal M}_n | \quad \text{almost surely.}
$$ 

The case of unimodular matrices (i.e. ${\mathcal A}_{n}\in \text{SL}(2,{\mathbb R})$) is of particular interest.
In this case, under the natural additional assumption of 
{\em noncompactness}\footnote{The support of the distribution of ${\mathcal A}_{1}$ is not contained 
in any compact subgroup 
of $\text{SL}(2,{\mathbb R})$.} and {\em strong irreducibility},\footnote{There is no finite
union of proper subspaces $V$ with
${\mathcal A}_{1}V=V$, for all realizations of ${\mathcal A}_{1}$.} Furstenberg's theorem 
(\cite{Fu}, cf. \cite{BoLa} p. 30), asserts that the 
Lyapunov exponent is {\em strictly positive}. 
Moreover,  there exists a unique, continuous, $\mu$-invariant measure $\nu$ on the projective
line $P \left ( {\mathbb R}^2 \right )$--- that is, a
measure that is invariant under the projective action of matrices drawn
from the distribution $\mu$ (cf. \cite{BoLa} p. 30).

The calculation of the Lyapunov exponent involves this $\mu$-invariant measure, but
there are remarkably few non-trivial cases where $\nu$ has been found explicitly; 
three well-known examples will be discussed presently
(others are given in \cite{CL}) . The prominent feature shared by these examples is that
the invariant measure is found by considering a random continued fraction 
derived from the projective action of the relevant matrix ensemble.

The first example dates back to Dyson's work on the disordered chain problem. 

\begin{example}[Dyson \cite{Dy}]
\label{dysonExample}
The disordered chain is modeled by a system of harmonic oscillators coupled 
by linear forces. A physical realisation is obtained  by considering a
sequence of $N$ particles joined by elastic springs
obeying Hooke's law.  Denote
the mass and the displacement of the $n$th particle from its equilibrium position by 
$m_n$ and $x_n$ respectively, and
let $k_n$ be the elastic modulus of the spring between the $n$th and $(n+1)$th
particle. Then the
equation of motion takes the form
$$
m_{n}\,\ddot{x}_{n}=k_{n}(x_{n+1}-x_{n})+k_{n-1}(x_{n-1}-x_{n}), \quad 1 \le n \le N\,.
$$
By introducing additional variables, it is straightforward to express this as the first-order system
$$
\dot{u}_{n}=\sqrt{a_{n}} \,u_{n+1} - \sqrt{a_{n-1}} \,u_{n-1}, \quad 1 \le n \le 2N-1,
$$
where 
$$
a_{2n-1}:=\frac{k_{n}}{m_{n}}, \quad 
a_{2n}:=\frac{k_{n}}{m_{n+1}}\,.
$$
In matrix notation, 
$$
\dot{\mathbf u}={\mathcal J} {\mathbf u},
$$
where ${\mathcal J}$ is the tridiagonal (Jacobi) matrix
$$
{\mathcal J}=\begin{pmatrix}
0 & \sqrt{a_{1}} & 0 & 0 & 0 & \hdots & 0 \\
-\sqrt{a_{1}} & 0 & \sqrt{a_{2}} & 0 &  0 & \hdots & 0 \\
0 & -\sqrt{a_{2}} & 0 & \sqrt{a_{3}} & 0 & \hdots & 0 \\
\vdots & & & & & & \vdots \\
0 & 0 & \hdots & 0 & -\sqrt{a_{2N-3}} & 0 & \sqrt{a_{2N-2}} \\
0 & 0 & \hdots & 0 & 0 & - \sqrt{a_{2N-2}} & 0 \\ 
\end{pmatrix}\,.
$$
Dyson studied the spectral problem for ${\mathcal J}$ when the $a_n$
are independent, identically distributed random variables. It turns out
that the spectral properties of ${\mathcal J}$ (e.g. the eigenvalue density function) can be deduced
from the so-called characteristic function of the chain
$$
\Omega(t) := 2\,\int_{0}^{\infty}\,\ln(1+x)\nu_t (\d x),
$$ 
where $\nu_t$ is the distribution of the random continued fraction
\begin{equation}
\cfrac{a_{1} t}{1 + \cfrac{a_{2} t}{
1+  \cfrac{a_3 t}{1+ \cdots}}}\,. 
\label{dysonContinuedFraction}               
\end{equation}
To illustrate his approach,
Dyson elaborated the particular case where the $a_{n}$ are gamma-distributed,
i.e. for every Lebesgue-measurable subset $S$ of ${\mathbb R}_+$,
$$
\text{Pr} \left ( a_n \in S \right ) = \int_S \gamma_{p,s} (a) \,\d a,
$$
where
\begin{equation}
\gamma_{p,s}(a) :=
\frac{1}{s^p \Gamma(p)} \,a^{p-1} \e^{- \frac{a}{s}}\,,
\quad a \ge 0; \;p,s >0\,.
\label{gammaDistribution}
\end{equation}
Then the probability density function of the random continued 
fraction (\ref{dysonContinuedFraction}) is given explicitly by
$$
\nu_t(\d x)=C_{p,s,t}\frac{x^{p-1}}{(1+x)^p} \exp \left (- \frac{x}{s t} \right )\,\d x,
$$
where $C_{p,s,t}$ is a normalisation constant. 

In terms of products of random matrices,
Dyson's continued fraction (\ref{dysonContinuedFraction}) corresponds to the case
where
$$
{\mathcal A}_n = \begin{pmatrix}
0 & a_n t \\
1 & 1
\end{pmatrix}
$$
in the product (\ref{productOfRandomMatrices}). The distribution $\nu_t$ of the continued fraction
is the $\mu$-invariant distribution associated with this product.
\end{example}

One of the present paper's contribution is the 
calculation of the distribution of a random continued fraction that arises
in another important physical model,
namely the discrete Schr\"odinger equation with a random potential.  
In this regard, Lloyd's model (cf. \cite{Ll}, \cite{HJ}, \cite{Th}) 
with a Cauchy-distributed potential in one spatial dimension 
is possibly the best-known example where the 
invariant measure and the corresponding Lyapunov exponent have been found explicitly. 

\begin{example}[Lloyd's model \cite{BoLa}, p. 35] 
\label{cauchyExample}
Let $z=x+ \i y \in {\mathbb C}$ with $y>0$ and denote by $C_{z}$ the distribution
of the random variable $x+y c$, where $c$ is Cauchy-distributed, i.e
$$
\text{Pr} \left ( c \in S \right ) = \frac{1}{\pi} \int_S \frac{\d t}{1+t^2}
$$ 
for every Lebesgue-measurable subset $S$ of ${\mathbb R}$.
Set
$$
{\mathcal A}_{n} :=
\begin{pmatrix}
a_{n} & -1 \\
1    &  0
\end{pmatrix} \in \text{SL}(2,{\mathbb R}),
$$
where the $a_{n}$ are independent and $C_{z}$-distributed. The invariant measure $\nu$
is then given by the distribution $C_{u}$, where $u$ and $z$ are
related by $z=u+1/u$. It follows easily that 
$$
\lambda=\ln|u|\,.
$$
\end{example}

We end our brief survey with an example that, once again,
involves a continued fraction with gamma-distributed elements;
the resulting invariant distribution is a so-called generalised inverse Gaussian distribution.

\begin{example}[The generalised inverse Gaussian distribution \cite{BoLa}, pp. 170-171]
\label{letacSeshadriExample}
Set
\begin{equation}
{\mathcal A}_n :=
\begin{pmatrix}
0 & 1 \\
1 & a_n 
\end{pmatrix},
\label{letacSeshadriMatrix}
\end{equation}
where the $a_n$ are independent, gamma-distributed random variables
with parameters $p$ and $s$.
It was shown by Letac and Seshadri \cite{LS1} that the probability density function of
the invariant measure $\nu$
is then
\begin{equation}
f_0(x) := \frac{1}{2 K_p (2/s)} \,x^{-p-1} \exp \left [ - \frac{1}{s} \left ( x + \frac{1}{x} \right ) \right ]
\,, \quad x \ge 0,
\label{letacSeshadriDistribution}
\end{equation}
where $K_p$ is the modified Bessel function of order $p$. As we will show below, the
Lyapunov exponent can be expressed in terms of modified Bessel functions. The details
can be found in Section \ref{lyapunov}. See also \cite {LS2} and 
\cite{Be} for some generalisations of this example. 
\end{example}

In the papers that form the basis of Examples \ref{dysonExample} and \ref{letacSeshadriExample}, 
the authors were concerned --- not with products of 
random matrices--- but rather with the problem of determining the distribution of a
continued fraction with random coefficients. We have already mentioned
Dyson's continued fraction (\ref{dysonContinuedFraction}). The continued
fraction studied by Letac \& Seshadri is  of the form
\begin{equation}
X = \cfrac{1}{a_1  + \cfrac{1}{
a_2 + \cfrac{1}{a_3 
+ \cdots}}}\,,
\label{letacSeshadriContinuedFraction}
\end{equation}
where, as in Dyson's case, the $a_{n}$ are independent and gamma-distributed.

In this paper, we generalise this example to the case where
the elements take values along a ray
in the complex plane. Hence, from now on, unless explicitly stated otherwise, we consider matrices 
in $\text{GL}(2,{\mathbb C})$.


\subsection{Main results} 


Fix a constant
$$
\alpha \in (-\pi/2,\pi/2)
$$
and consider the one-parameter family of complex matrices of the form
\begin{equation}
{\mathcal A}_n (\alpha) :=
\begin{pmatrix}
0 & 1 \\
1 & a_n \e^{\i \alpha} 
\end{pmatrix},
\label{randomMatrixForAlpha}
\end{equation}
where the $a_n$ are, again, independent gamma-distributed random variables. 

The corresponding random continued fraction is
\begin{equation}
Z = \cfrac{1}{a_1 \e^{\i \alpha} + \cfrac{1}{
a_2 \e^{\i \alpha}+ \cfrac{1}{a_3 \e^{\i \alpha}
+ \cdots}}}\,.                
\label{continuedFraction}
\end{equation}
The random variable $Z$ takes values in the cone
\begin{equation}
S_{\alpha} = \left \{ z \in {\mathbb C}: \; |\arg z| \le |\alpha | \right \}\,.
\label{theSetS}
\end{equation}

\begin{figure}[htbp]
\vspace{15cm} 
\begin{picture}(0,0) 
\put(-95,210){(a)}
\put(100,210){(b)} 
\put(-95,0){(c)}
\put(100,0){(d)} 
\end{picture} 
\includegraphics{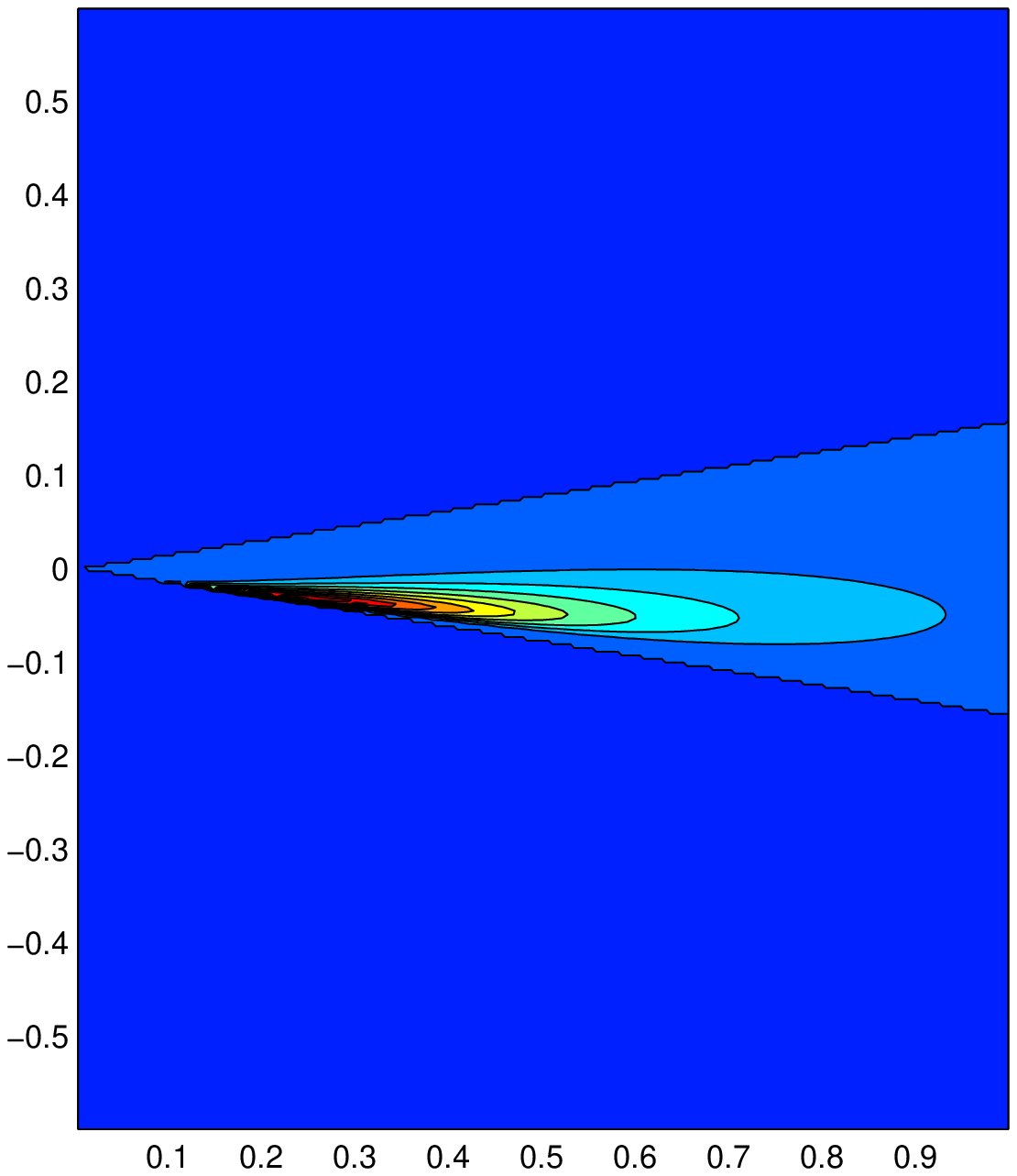}  
\includegraphics{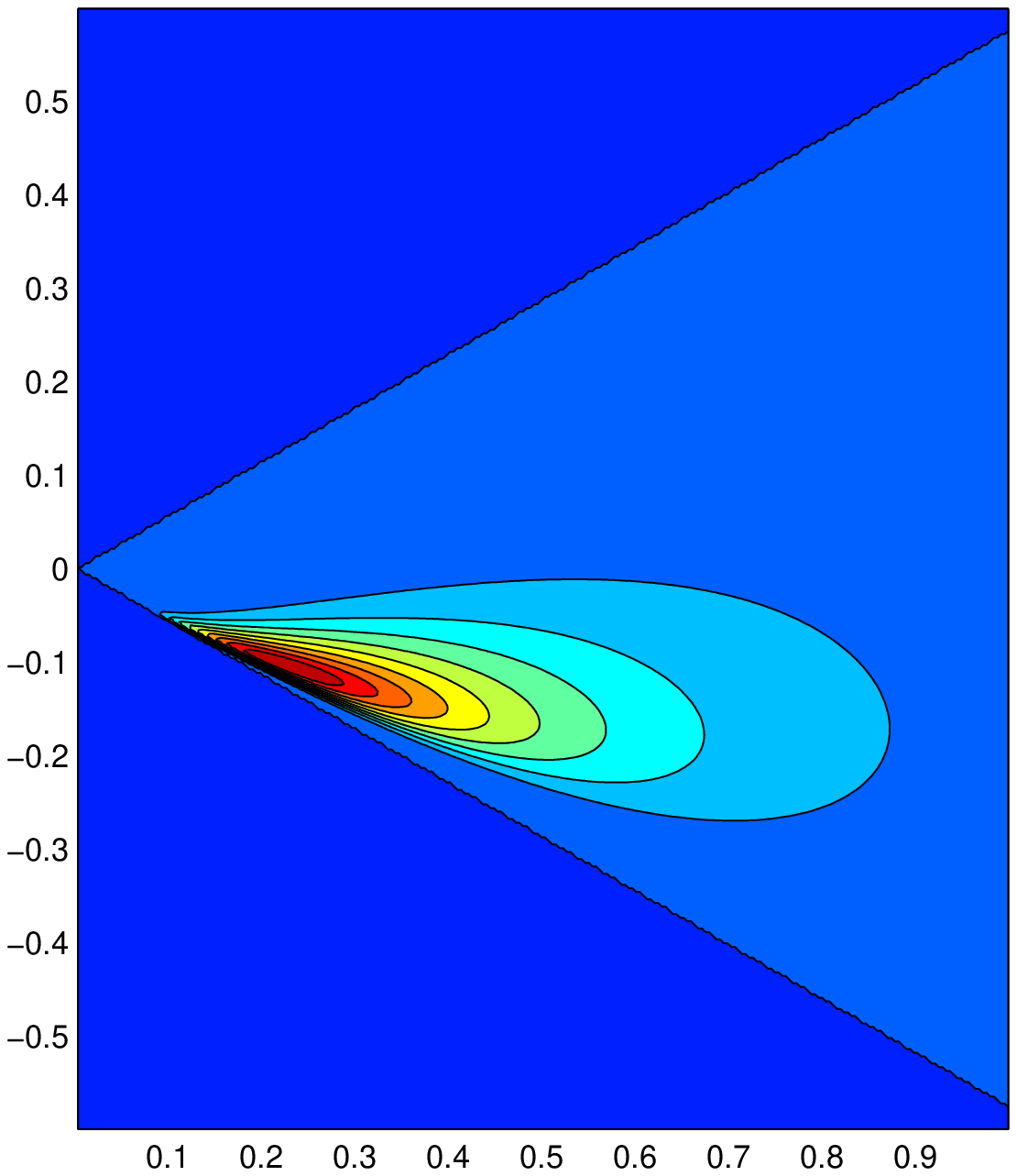} 
\includegraphics{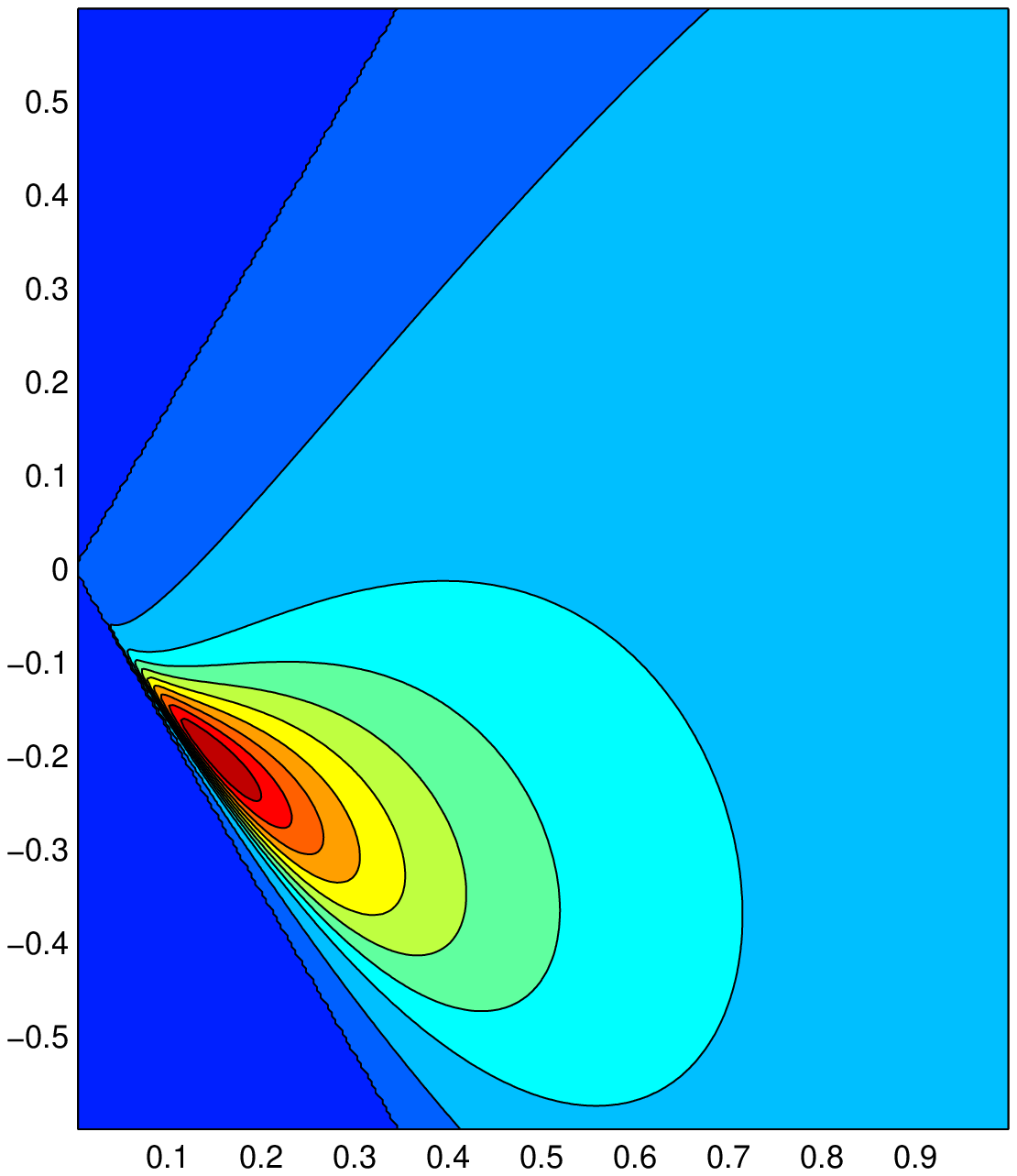}  
\includegraphics{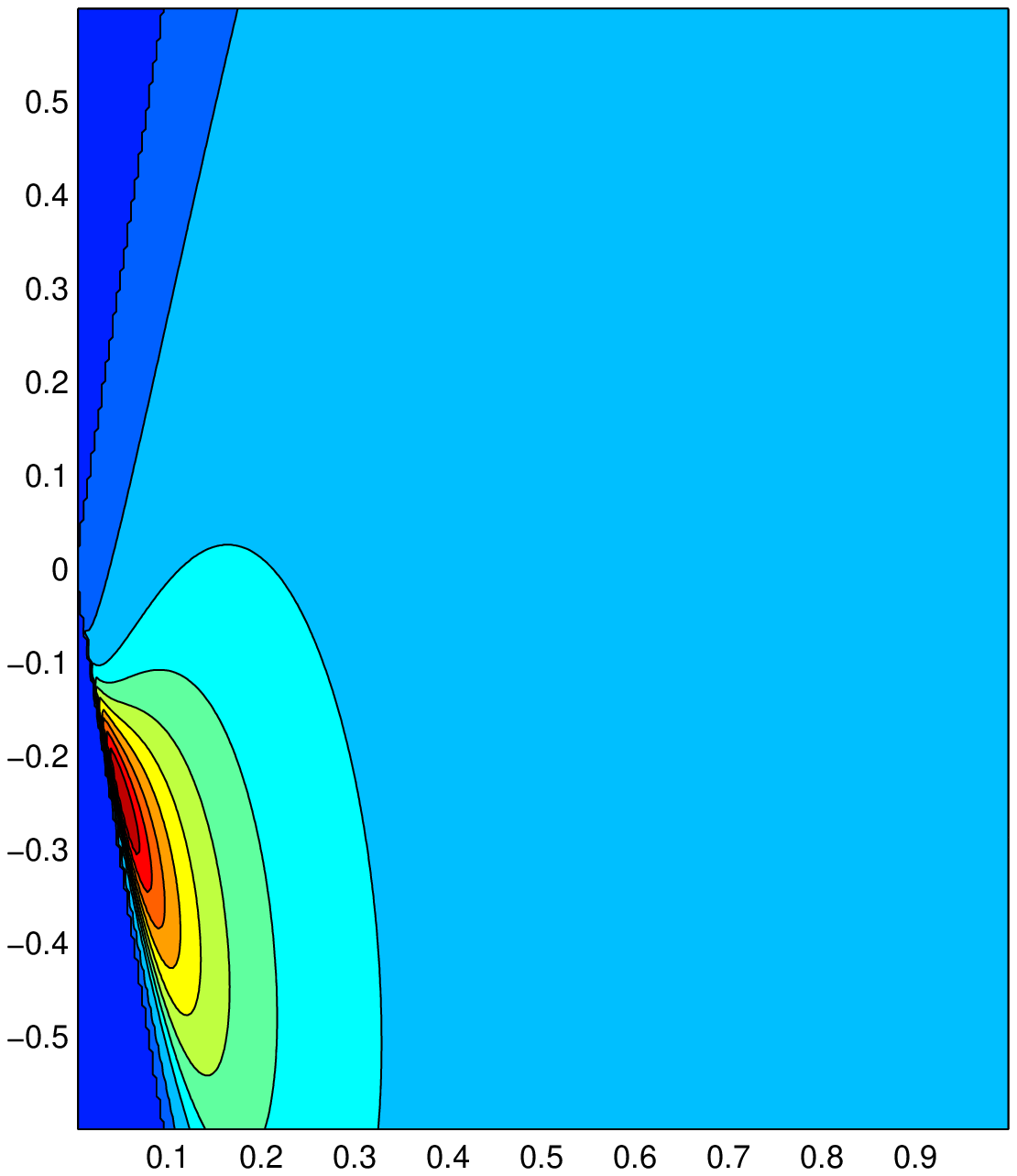} 
\caption{The probability density function
of $Z$ for $p=s=1$ and four different values of $\alpha$: (a) $\pi/20$, (b) $\pi/6$, (c)
$\pi/3$ and (d) $9\pi/20$. Blue and red correspond to low and high probability respectively.
For values of $\alpha$ close to $0$ (resp. $\pi/2$) the density is localised near the positive real half-line (resp. the imaginary axis), see Theorem \ref{alphaIsPiOver2Theorem} and \S \ref{weak}.}
\label{densityFigure} 
\end{figure}

Our main contribution is an explicit formula for the distribution of $Z$ or,
equivalently, for the $\mu$-invariant measure $\nu$ associated with the infinite product
of the random matrices (\ref{randomMatrixForAlpha}).
\begin{theorem}
Let $0 < |\alpha | < \pi/2$. Suppose that the $a_n$ are independent, gamma-distributed random variables
with parameters $p,s>0$, and  write $z = r \text{\em e}^{\text{\em i} \theta}$.
Then the probability density function of $Z$--- equivalently, that
of the $\mu$-invariant measure $\nu$--- is supported on $S_{\alpha}$ and given by
\begin{multline}
\label{densityFunction}
f_\alpha (z) = 
\frac{\sin(2 |\alpha|)}{\left | 2 K_p \left ( 2 \e^{\text{\em i} \alpha}/s \right ) \right |^{2}} 
\frac{1 }{r^2 \sin^2(\alpha+\theta)}
\left [ \frac{\sin(\alpha-\theta)}{\sin(\alpha+\theta)}
\right ]^{p-1} \\
\times \exp \left \{ -\frac{\sin(2 \alpha)}{s}
\left [ \frac{1}{r
\sin(\alpha-\theta)}+ \frac{r}{\sin(\alpha+\theta)} \right ] \right \}\,.
\end{multline}
In particular, the density is a smooth function that decays exponentially fast at infinity in every 
direction contained in the cone $S_{\alpha}$.
\label{generalisedLetacSeshadriThm}
\end{theorem}
Plots of the probability density function $f_{\alpha}$ for $p=s=1$
and various values of $\alpha$ are shown in Figure \ref{densityFigure}.

In Section \ref{lyapunov}, we use the above result to calculate the corresponding Lyapunov exponent and 
express it in terms of the logarithmic derivative of the modified Bessel function, namely
(see Theorem \ref{LyEx})
\begin{equation}
\notag
\lambda_{p,s}(\alpha)=\Real
\frac{\partial_{p} K_{p}\left(\frac{2}{s} e^{i\alpha}\right)}{K_{p}\left(\frac{2}{s} e^{i\alpha}\right)}.
\end{equation}

By considering the weak limit in (\ref{densityFunction}) as $\alpha \rightarrow 0$, we recover the distribution
found originally by Letac and Seshadri (see Section \ref{weak}). 

\begin{figure}[htbp]
\vspace{7.5cm} 
\begin{picture}(0,0) 
\put(-75,115){$\alpha=\frac{1}{3}\pi$}
\put(-98,190){$\alpha=\frac{31}{64}\pi$}
\put(-10,190){$\alpha=\frac{1}{6}\pi$}
\put(0,10){$\theta$}
\end{picture} 
\includegraphics{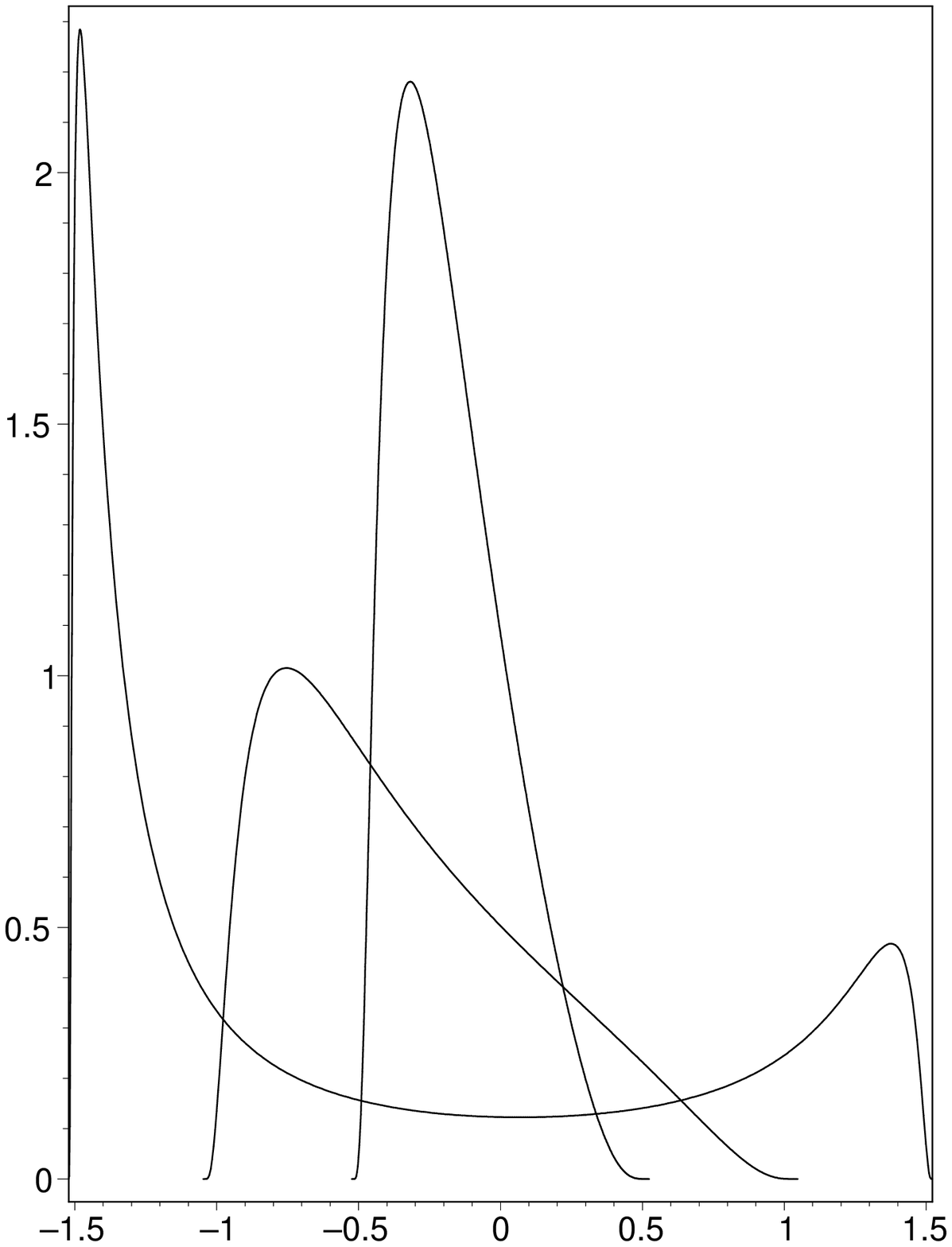}  
\caption{The probability density function
of $\arg Z$ for $p=s=1$ and various values
of $\alpha$.}
\label{argumentFigure} 
\end{figure}

\begin{figure}[htbp]
\vspace{7.5cm} 
\begin{picture}(0,0) 
\put(-128,124){\small{{\em s}=1/4}}
\put(-132,142){\small{{\em s}=1/2}}
\put(-135,195){\small{{\em s}=1}}
\put(-100,0){(a)}
\put(140,40){\small{{\em p}=1}}
\put(76,118){\small{{\em p}=2}}
\put(60,150){\small{{\em p}=3}}
\put(100,0){(b)} 
\end{picture} 
\includegraphics{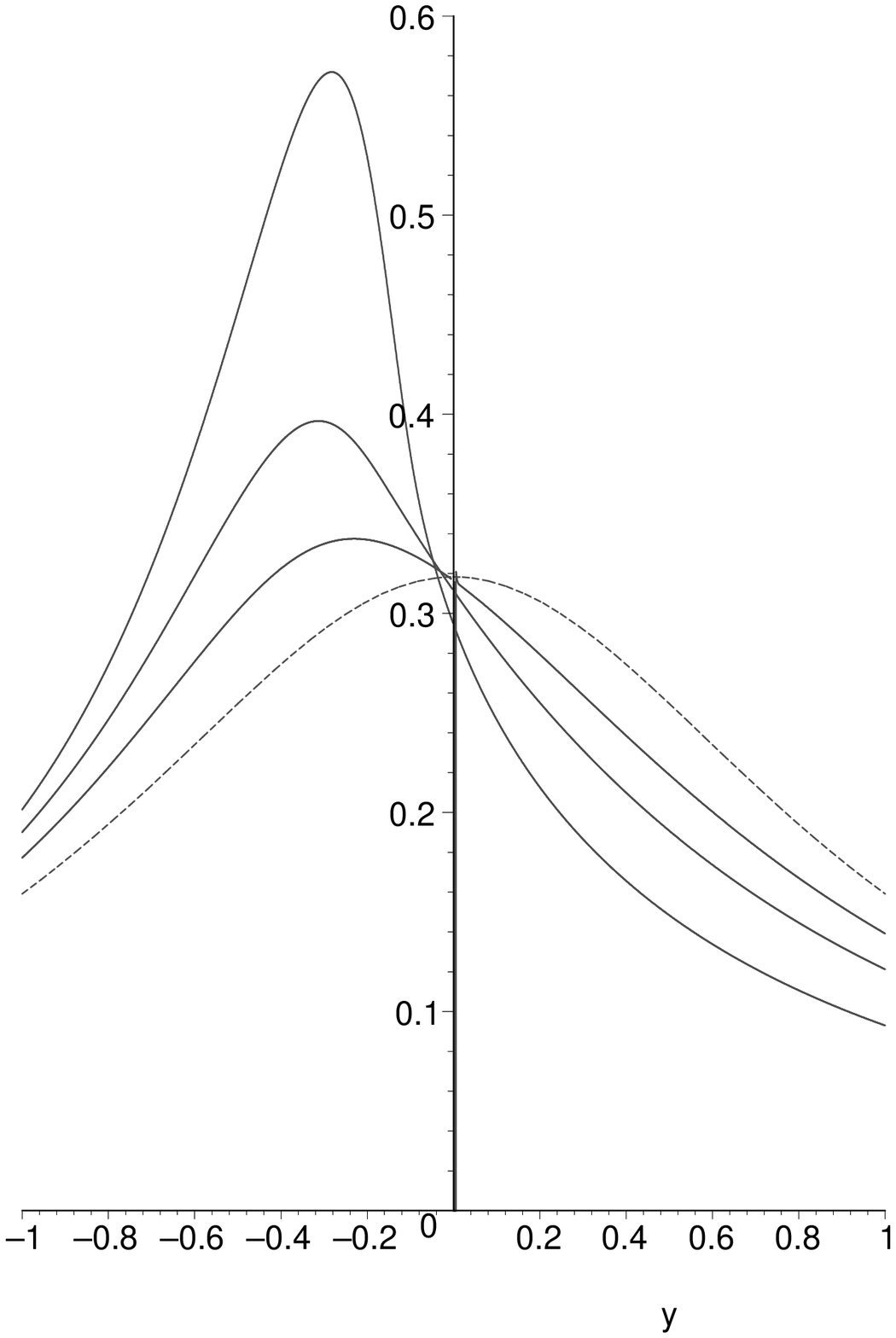}  
\includegraphics{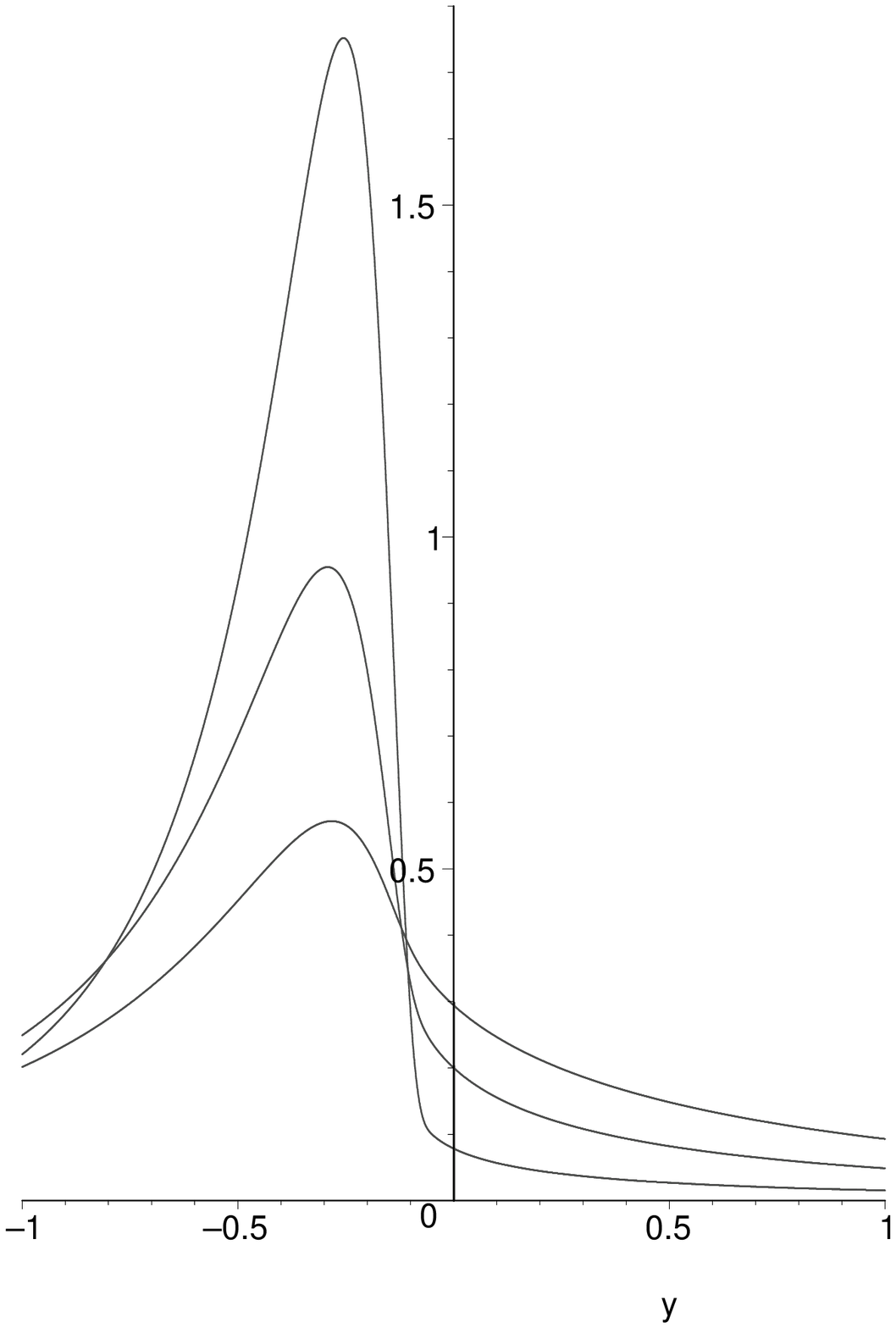} 
\caption{The probability density function 
$f_{\pi/2}$ for $x=0$ and
various values of the parameters $s$ and $p$: (a) $p=1$; (b) $s=1$. The dashed curve
in (a) corresponds to the Cauchy probability density function $\frac{1}{\pi} \frac{1}{1+y^2}$.}
\label{piOver2PdfFigure} 
\end{figure}

The behaviour of the random variable $Z$ as $|\alpha| \rightarrow \pi/2-$ is particularly interesting.
Figure \ref{argumentFigure} shows plots of the probability density function of $\arg Z$ for various
values of $\alpha$ when $p=s=1$.
As $|\alpha|$ approaches $\pi/2$,
the support of the measure $\nu$ becomes concentrated on the imaginary
axis.
The weak limit as $|\alpha |\rightarrow \pi/2-$ leads to our second result: 

\begin{theorem}
Let $|\alpha | = \pi/2$. Suppose that the $a_n$ are independent and gamma-distributed,
with parameters $p,s>0$, and write $z=x + \text{\em i} y$. Then,
the probability density function of $Z$--- equivalently, that of the $\mu$-invariant
distribution $\nu$--- is given by
\begin{multline}
\notag
f_{\pm \pi/2}(z) = 
\frac{\delta(x)}{\pi^2 \left [ J_p^2(2/s)+Y_p^2(2/s) \right ]} \,\frac{1}{y^{p+1}} 
\exp \left [ \pm \frac{1}{s} 
\left ( \frac{1}{y} -y \right ) \right ] \\
\times \int_{c(y)}^y 
\exp \left [ \mp \frac{1}{s} 
\left ( \frac{1}{t} -t\right ) \right ]\,t^{p-1}\,\d t\,,
\end{multline} 
where
$$
c(y) = \begin{cases}
-\infty & \text{if $y<0$} \\
0 & \text{if $y>0$}
\end{cases}
$$
and $\delta$ is the Dirac delta. In particular, $\nu$ 
is supported on the imaginary axis, where
the density decays algebraically (like $1/y^{2}$) at $\pm \infty$.
\label{alphaIsPiOver2Theorem}
\end{theorem}
Plots of $f_{\pi/2}(\i y)$ for various values of $p$ and $s$ 
are shown in Figure \ref{piOver2PdfFigure}. For comparison, the figure includes a
plot of the Cauchy probability density function. We shall see in \S \ref{alphaIsPiOver2}
that, for $\alpha = \pm \pi/2$ and $s$ small, $Z$ is approximately Cauchy-distributed along the imaginary
axis.

Next, we discuss some interesting applications of these results.


\subsection{Random Schr\"odinger operators} 


In 1958, P. W. Anderson \cite{An} postulated that
the spectra of Schr\"odinger operators with random potentials should exhibit
a tendency towards a dense pure point spectrum with bound eigenstates. When the 
eigenstates decay exponentially fast, one speaks of strong or exponential localisation. 
This type of localisation has been proved for one-dimensional
systems; 
cf. \cite{CaLa}, Chap. VIII. Higher-dimensional systems will not be considered here; we
refer the reader to \cite{CaLa} for a discussion of that case.

In order to make the connection between this general theory and our own results,  
let us first note that by
conjugating the matrices (\ref{randomMatrixForAlpha}) for $\alpha=\pi/2$ with a rotation that
maps $P( {\mathbb R}^2)$ into the imaginary line, we obtain 
$$
R^*{\mathcal A}_n (\alpha) R=
\i \begin{pmatrix}
0 & -1 \\
1 & a_n  
\end{pmatrix},
\qquad \text{ where } \quad 
R=
\begin{pmatrix}
\i & 0 \\
0 & 1  
\end{pmatrix}.
$$
Thus the corresponding random matrices reduce in this 
limiting case to the transfer matrices for the discrete stationary one-dimensional 
Schr\"odinger equation 
\begin{equation}
(-\Delta +V)\,y= E\, y\,
\label{discreteSchroedingerEquation}
\end{equation}
with $y=\{y_{n}\}_{n\in\mathbb{Z}}$, $(\Delta y)_{n}= y_{n+1}-2y_{n}+y_{n-1}$ (tight-binding, nearest-neighbour laplacian), 
 $(V\,y)_{n}=a_{n}y_{n}$ (independent identically distributed random potential) and energy level
$E=2$ (cf. \cite{BoLa}, p. 187). 

It is readily seen that the above equation can be rewritten in the form
\begin{equation}
\notag
\begin{pmatrix}
y_{n} & y_{n+1}
\end{pmatrix}
= 
\begin{pmatrix}
y_{n-1}  & y_{n}
\end{pmatrix}
\begin{pmatrix}
0 & -1 \\
1 & a_n
\end{pmatrix}
= 
(-\i)^n\,\begin{pmatrix}
y_{0} & y_{1}
\end{pmatrix}
R^\ast
{\mathcal M}_n(\pi/2) R\,.
\label{productOfMatrices}
\end{equation}

It is well known \cite{KS} that the spectrum on $\ell^{2}(\mathbb{Z})$ of the Hamiltonian
associated with Equation (\ref{discreteSchroedingerEquation}), i.e of the operator 
$H=-\Delta +V$ 
is given, with probability one, by
\begin{equation}
\sigma(H)=\sigma(-\Delta)+\supp(\mu),
\end{equation}
where $\sigma(-\Delta)=[0,4]$ and $\mu$ denotes the probability measure of the potential.

It has been shown in \cite{CKM} that, as long as $\mu$ is
not concentrated on a single point and possesses
some finite moment, the spectrum $\sigma(H)$ is pure-point with
probability one. The eigenfunctions are exponentially
localised with a rate determined by the {\em value} of the Lyapunov exponent (see \cite{CaLa,BoLa}).
Thus is view of our results we have
\begin{proposition}
Let $H$ be the hamiltonian of the Anderson tight-binding model in dimension 1 with 
$\gamma_{p,s}$-distributed potential.
The exponential localization rate for the energy level
$$
E=2\in\sigma(H)=(0,+\infty)
$$
is given by
$$
\lambda_{s,p}(\pi/2)=\Real\frac{\partial_{p}K_{p}(2i/s)}{K_{p}(2i/s)}.
$$
\end{proposition}

We note that the Lyapunov exponent is a smooth 
function of the energy level $E$ (cf. \cite{CaLa}, Proposition VIII.1.1).
Thus, although the above result is not sufficient  to give estimates of the decay rate of all the eigenstates,
it does yield quantitative 
information in the close enough neighbourhood of $E=2$, which is almost surely in the bulk of the spectrum.


\subsection{Random Stieltjes functions}


Another motivation for this work is to study the convergence of Pad\'{e}
approximation for series that are random in some sense \cite{TD}.
Consider
the class of random Stieltjes functions whose continued fraction
expansion is 
\begin{equation}
\notag
F(t) = \cfrac{1}{c_1+\cfrac{t}{c_2 +\cfrac{t}{c_3 + \cdots}}}\,, \quad t \in {\mathbb C} \backslash
{\mathbb R}_-,
\end{equation}
where the $c_n$ are independent draws from the gamma distribution with parameters
$p$ and $\sigma$. By 
truncating this continued fraction, we obtain  rational approximations of $F$:
\begin{equation}
F_n (t) := 
\cfrac{1}{c_1+\cfrac{t}{c_2 +  \cdots + \cfrac{t}{c_{n}}}}\,, \quad t \in {\mathbb C} \backslash
{\mathbb R}_-\,.
\label{stieltjesConvergent}
\end{equation}
It is a well-known fact that the $F_n$ are diagonal or near-diagonal Pad\'{e} approximants of $F$
\cite{BGM,BO}. Roughly speaking, this means that 
$$
F(t) - F_n(t) = O(t^n) \quad \text{as $t \rightarrow 0$},
$$
and that there is no other rational function--- with a numerator and a denominator of 
lesser degree--- with this property.
The rate of decay of the error of Pad\'{e} approximation as $n \rightarrow \infty$, $t$ fixed, is
of considerable practical interest.

A simple calculation
reveals that the distribution of $\sqrt{t} F(t)$ is precisely that of the random variable $Z$ defined
by Equation (\ref{continuedFraction}) if we take
$$
a_n = \frac{c_n}{\sqrt{|t|}}, \quad s = \frac{\sigma}{\sqrt{|t|}} \quad \text{and}
\quad \alpha = -\frac{\arg t}{2}\,.
$$
The rate of convergence of
Pad\'{e} approximation for a typical realisation of the function $F$ is then given by
$$
\frac{\ln \left | F(t)-F_n(t) \right |}{n}
\xrightarrow[n \rightarrow \infty]{} 
-2 \lambda_{p,\frac{\sigma}{\sqrt{|t|}}} \left ( - \frac{\arg t}{2} \right )\quad \text{almost surely.}
$$
The proof of this result requires 
a careful study of the ergodic properties of the Markov
chain associated with the continued fraction; 
a detailed treatment will be given in  a separate publication.

The remainder of the paper provides the details of
the proofs of these assertions. In the next section,
we introduce some notation, elaborate the correspondence between
products of $2 \times 2$ matrices and continued fractions,  and derive an integral equation for the
unknown probability density function of $Z$.
In \S \ref{alphaSection}, we deduce
a partial differential
equation, and solve it.
The case $\alpha = \pm \pi/2$ is treated
in \S \ref{alphaIsPiOver2} and, in \S \ref{weak}, we show that the singular measure obtained in this
case is a weak limit of that found earlier.
Finally,
\S \ref{lyapunov} is concerned with the explicit
calculation of the Lyapunov exponent.

\section{Preliminaries}
\label{prelim}

The correspondence between continued fractions and products of $2 \times 2$ matrices is well-known
(see for instance \cite{BoLa}, Part A, \S VI.5), but since it forms the basis of our approach, it is
desirable to elaborate it here.
The starting point is the observation that 
any linear fractional transformation 
$$
{\mathcal F}(z) = \frac{a_{11}z + a_{12}}{a_{21} z + a_{22}}
$$ 
corresponds to the action of a matrix
\begin{equation}
{\mathcal A} = \begin{pmatrix}
a_{11} & a_{12} \\
a_{21} & a_{22}
\end{pmatrix}
\label{matrixDefn}
\end{equation}
on a complex or real projective line $P({\mathbb C}^{2})$.
Indeed, denoting by 
$$
{\mathbf z} = \begin{pmatrix} 
z_{1} \\
z_{2}
\end{pmatrix}
$$ 
any nonzero vector in ${\mathbb C}^{2}$ and defining its 
projection by
\begin{equation}
\notag
\Pi {\mathbf z} =
\begin{cases}
\frac{z_{1}}{z_{2}} & \text{if $z_{2}\not= 0$} \\
\infty & \text{otherwise} 
\end{cases},
\end{equation}
we obtain a natural identification of $P({\mathbb C}^{2})$ 
with $\Pi {\mathbb C}^{2} =\overline{{\mathbb C}}={\mathbb C} \cup \{\infty\}$.
The action of ${\mathcal A}$ on $P({\mathbb C}^{2})$ (denoted by $\cdot$) 
can then be defined via
\begin{equation}
{\mathcal A} \cdot z = \Pi {\mathcal A} \begin{pmatrix}
z \\
1
\end{pmatrix}
 = \frac{a_{11}z+a_{12}}{a_{21}z+a_{22}}
= {\mathcal F}(z)\,.
\end{equation}

Then, for the choice
\begin{equation}
{\mathcal A} = {\mathcal A}(\alpha) := \begin{pmatrix}
0 & 1 \\
1 & a \e^{\i \alpha}
\end{pmatrix},
\quad |\alpha | \le  \frac{\pi}{2},
\label{randomMatrix}
\end{equation}
we get
\begin{equation}
{\mathcal A} \cdot z = \frac{1}{z+a \e^{\i \alpha}}=: {\mathcal F}_{a}(z)\,.
\label{linearFractionalTransformation}
\end{equation}

It is convenient to extend
the notation for the set $S_{\alpha}$, 
defined by (\ref{theSetS}) for  $0 <|\alpha| < \pi/2$,
to the limiting cases $\alpha = 0, \pm \pi/2$. We shall
write
$$
S_{0}:={\mathbb R}_{+} \quad \text{and} \quad
S_{-\pi/2}=S_{\pi/2}:=\i {\mathbb R} =\{z\in {\mathbb C} : |\arg z| = \pi/2 \}\,.
$$
Then, for every $a \ge 0$,
$$
{\mathcal F}_a \left ( S_\alpha \right ) \subseteq S_\alpha\,.
$$
Now, suppose that $a$ is a positive random variable distributed according to
a probability measure $\mu (\d a)$ on ${\mathbb R}_+$, and let $\hat{Z}_{0}$ be
an arbitrary random variable taking values in $S_\alpha$. 
Consider the iteration
\begin{equation}
\hat{Z}_{n} = {\mathcal F}_{a_n} \left (\hat{Z}_{n-1} \right ), 
\;n \in {\mathbb N},
\label{forwardIteration}
\end{equation}
where the $a_n$ are independent draws
from $\left ({\mathbb R}_+,\mu \right)$.
Under appropriate conditions on $\mu$ \cite{DF}, it can be shown
that the Markov chain defined by Equation (\ref{forwardIteration})
has a stationary distribution $\nu(\d z)$ independent
of the starting value $\hat{Z}_0$. Moreover, the latter distribution coincides with 
the distribution of the random continued fraction  (\ref{continuedFraction}), which
can be constructed by means of the successive applications of the corresponding
random matrices 
\begin{multline}
\notag
{\mathcal M}_n (\alpha) \cdot z :=  {\mathcal A}_1 (\alpha) {\mathcal A}_2(\alpha) \cdots {\mathcal A}_n (\alpha) \cdot z = 
{\mathcal F}_{a_1} \circ {\mathcal F}_{a_2} \circ
\cdots \circ {\mathcal F}_{a_n} (z)\\
\\ = \cfrac{1}{a_1 \e^{\i \alpha} +\cfrac{1}{a_{2} \e^{\i \alpha}+\cdots+\cfrac{1}{a_n \e^{\i \alpha}+z}}}\,.
\end{multline}
Hence, the $\mu$-invariant measure on the projective line associated with the infinite
product of the ${\mathcal A}_n$ is precisely $\nu(\d z)$.

Let us denote by $\nu_n$ the distribution
of $\hat{Z}_n$.
We are only interested in the
case where the measures $\nu_n(\d z)$
and $\nu(\d z)$
on $S_\alpha$ are absolutely continuous
with respect to the Lebesgue measure $\d z$;
so let $f_{\alpha,n}$ and $f_\alpha$ be the
probability density functions of $\hat{Z}_n$
and $Z$ respectively, i.e.
$$
\nu(\d z) = f_\alpha(z) \d z \quad \text{and} \quad
\nu_n(\d z) = f_{\alpha,n}(z) \d z 
 \quad \text{for $z \in S_\alpha$}\,.
$$
Our aim is to find an explicit formula for $f_\alpha$.
To this end, we derive a recurrence relation for the $f_{\alpha,n}$
and hence, by taking
the limit as $n \rightarrow \infty$, find an integral
equation satisfied by $f_\alpha$.
We shall assume that $f_{\alpha,n}$ and $f_\alpha$ are smooth.
We introduce the map ${\mathcal G}_{a} : S_\alpha \rightarrow {\mathbb K}
:= {\mathbb R}$, $\i {\mathbb R}$ or ${\mathbb C}$
defined by
$$
{\mathcal G}_{a}(z) = \frac{1}{z}-a \e^{\i \alpha}\,.
$$
Thus, ${\mathcal G}_{a}$ is the left inverse of ${\mathcal F}_a$:
$$
\forall \, z \in S_\alpha, \quad 
\left ( {\mathcal G}_a \circ {\mathcal F}_{a} \right ) (z) = z\,.
$$
On the other hand, for $z \in S_\alpha$,
$$
\left ( {\mathcal F}_a \circ {\mathcal G}_{a} \right ) (z) = z
$$
only if ${\mathcal G}_{a}(z) \in S_\alpha$.

Let $S$ be a measurable set. We have
\begin{multline}
\notag
\Pr \left ( \hat{Z}_{n} \in S \right ) 
= \int_{{\mathbb R}_+} \Pr \left ( \hat{Z}_{n} \in S \,\left | \right . 
\,a_n = a \right )\,
\mu(\d a) \\
= \int_{{\mathbb R}_+} \Pr \left ( {\mathcal F}_{a} (\hat{Z}_{n-1}) \in S \right ) 
\,\mu(\d a)  =
\int_{{\mathbb R}_+} \Pr \left ( \hat{Z}_{n-1} \in {\mathcal G}_{a} (S) 
\cap S_\alpha \right ) 
\,\mu(\d a) \\
= \int_{A} \int_{{\mathcal G}_{a}(S) \cap S_\alpha} f_{\alpha,{n-1}}(z)\text{d}z 
\,\mu(\d a)
\end{multline}
where
$$
A := \left \{ a \in {\mathbb R}_+ : \; {\mathcal G}_a (S) \cap S_\alpha \ne
\emptyset
\right \}\,.
$$
For $a \in A$, we have
$$
\int_{{\mathcal G}_{a}(S) \cap S_\alpha} f_{\alpha,n-1}(z) \text{d} z
= \int_{{\mathcal F}_{a} \left ( {\mathcal G}_{a} (S) \cap S_\alpha \right )}
f_{\alpha,n-1} \left ( {\mathcal G}_{a} (w) \right ) \, {\mathcal J}_{a} (w)\,
\text{d} w,
$$
where ${\mathcal J}_{a}$ is the jacobian of the transformation
$$
z = {\mathcal G}_{a}(w)\,.
$$
Hence
$$
\Pr \left ( \hat{Z}_{n} \in S \right ) =
\int_{A} \int_{{\mathcal F}_{a} \left ( {\mathcal G}_{a}(S) \cap S_\alpha \right )}
f_{\alpha,n-1}\left ( {\mathcal G}_{a}(w) \right )\,{\mathcal J}_{a}(w)
\,\text{d}w \,\mu(\d a)\,.
$$
Now, consider the region 
$$
R := \left \{ (a,w) \in {\mathbb R}_+ \times S_\alpha: \;a \in A \;\;
\text{and} \;\; w \in {\mathcal F}_{a} \left ( {\mathcal G}_{a}(S) 
\cap S_\alpha 
\right ) \right \}\,.
$$
It is readily verified that
$$
R = \left \{ (a,w) \in {\mathbb R}_+ \times S_\alpha:\;
w \in S \;\; \text{and} \;\; a \in A(w) \right \},
$$
where
\begin{equation}
A(z) := \left \{ a \in {\mathbb R}_+ : \; 
{\mathcal G}_{a}(z) \in S_\alpha \right \}\,.
\label{equation4A}
\end{equation}
Hence, 
$$
\Pr \left ( \hat{Z}_{n} \in S \right ) =
\int_{S} \int_{A(w)}
f_{\alpha,n-1}\left ( {\mathcal G}_{a}(w) \right )\, {\mathcal J}_{a}(w) \,\mu(\d a) 
\,\text{d} w\,.
$$
It follows that
\begin{equation}
f_{\alpha,n}(z) = \int_{A(z)} f_{\alpha,n-1} \left ( {\mathcal G}_{a} (z) \right )\,
{\mathcal J}_{a}(z)\,\mu(\d a)\,.
\label{recurrenceForMu}
\end{equation}

Taking the limit as $n \rightarrow \infty$,
we conclude that, if the continued fraction (\ref{continuedFraction})
has a smooth probability density function $f_\alpha$, then 
$f_\alpha$ satisfies the equation
\begin{equation}
f_\alpha(z) = 
\int_{A(z)} f_\alpha \left ( {\mathcal G}_a(z) \right )\,
{\mathcal J}_{a}(z)\,\mu(\d a) \quad \forall \, z \in S_\alpha\,.
\label{equation4f}
\end{equation}

We shall henceforth assume that 
$$
\mu (\d a) = \gamma_{p,s}(a) \d a\,.
$$
The Lyapunov exponent will be denoted by
\begin{equation}
\lambda_{p,s} (\alpha) := 
\lim_{n \rightarrow \infty} \frac{1}{n} \ln
\left | {\mathcal M}_n (\alpha) \right |\,.
\label{upperLyapunovExponent}
\end{equation}
Here,
$$
\left | {\mathcal M} \right | = \sup_{\left | {\mathbf z} \right | =1}
\left | {\mathcal M} {\mathbf z} \right |\,,
$$
where $| \cdot |$ denotes the standard euclidean
norm on ${\mathbb C}^2$. We can express the Lyapunov exponent in terms of the measure $\nu$ as follows

\begin{equation}
\lambda_{p,s} (\alpha) = -  
\int_{S_{\alpha}}\ln |z|  \,\nu( \d z) = -\int_{S_\alpha} \ln |z| \,f_\alpha (z) \d z\,.
\label{lyapunovIntegral}
\end{equation}
A proof is given Appendix \ref{Lyap}.

\begin{remark}
An equivalent way of presenting our results is to work with the reciprocal of the continued
fraction (\ref{continuedFraction}), i.e.
\begin{equation}
a_1 \e^{\i \alpha} + \cfrac{1}{a_2 \e^{\i \alpha} + \cfrac{1}{a_3 \e^{\i \alpha}+ \cdots}}\,.
\label{reciprocalContinuedFraction}
\end{equation}
The corresponding linear fractional transformation is
$$
z \mapsto a \e^{\i \alpha} + 1/z
$$
and so the matrices in the product (\ref{productOfRandomMatrices}) are of the form
\begin{equation}
\label{umatrix}
\begin{pmatrix}
a \e^{\i \alpha} & 1 \\
1 & 0
\end{pmatrix}
\,.
\end{equation}
In particular, when $\alpha=\pi/2$, we have
$$
(-\i)\, R^\ast \begin{pmatrix}
a \e^{\i \alpha} & 1 \\
1 & 0
\end{pmatrix} R =
\begin{pmatrix}
a_n & -1 \\
1 & 0
\end{pmatrix}
\,.
$$
Readers familiar with the application of the theory of products of random matrices to the spectral theory of Schr\"odinger operators 
will easily recognize the Schr\"odinger transfer matrix on the right-hand side of this equation.

Finally, we remark that, if $0 < \alpha < \pi/2$
and $f_\alpha$ is the probability density function of the random continued
fraction (\ref{continuedFraction}), then the probability density function, say $u$, of 
(\ref{reciprocalContinuedFraction}) is given by
$$
u(z) := \frac{1}{|z|^4} f_\alpha(1/z)\,.
$$
It will sometimes be convenient to work with $u$ instead of $f_\alpha$. Note that in this case, i.e. for the matrices
of the form (\ref{umatrix}) the Lyapunov exponent
$\lambda_{p,s}$ takes the form (see Appendix \ref{Lyap})
\begin{equation}
\label{uLE}
\lambda_{p,s}(\alpha) = \int_{S_\alpha} \ln |z| \,u(z) \d z\,.
\end{equation}
\label{inverseRemark}
\end{remark}

Letac \& Seshadri \cite{LS1} considered the particular case $\alpha=0$.
By using the Laplace transform, they showed that
the density function of the stationary
distribution is given by
Equation (\ref{letacSeshadriDistribution}).
In this paper, we study the case $0<\alpha \le \pi/2$ by a different method.
Our approach is inspired by Alain Comtet's 
derivation of (\ref{letacSeshadriDistribution}) in the particular case
$\alpha=0$ and $p=1$ \cite{Co}: By using the fact that $\gamma_{1,s}$ is 
a simple exponential, he showed how to replace Equation
(\ref{equation4f}) by one that involves $f_0(1/x)$, $f_0(x)$
and a derivative of $f_0$.
We shall develop this idea
and seek to derive a differential equation for
the unknown density function $f_\alpha$, making use of  the fact that,
for $p \in {\mathbb N}$,
the probability density function
$\gamma_{p,s}$ of the gamma distribution solves the differential
equation
\begin{equation}
\left [ \frac{\d}{\d a} + \frac{1}{s} \right ]^p
\gamma_{p,s} = 0, \quad
\frac{\d^{k}\gamma_{p,s}}{\d a^{k}}(0) =
\begin{cases}
0 & \text{if $0 \le k < p-1$} \\
\frac{1}{s^p} & \text{if $k=p-1$}
\end{cases}
\,.
\label{ode4gamma}
\end{equation}
The key observation is that this
is a linear equation with
constant coefficients.

\section{The case $0 < | \alpha | < \pi/2$}
\label{alphaSection}
Equation (\ref{equation4f}) states that
\begin{equation}
f_\alpha(z) = \frac{1}{|z|^4} \int_0^{a(1/z)} f_\alpha \left ( \frac{1}{z} -
a \e^{\i \alpha} \right )\,\mu(\d a),
\quad z \in S_\alpha,
\label{equation4fWhenAlphaIsNotZero}
\end{equation}
where
\begin{equation}
a(z) = \frac{|z| \sin (\alpha+\arg z)}{\sin(2 \alpha)}\,.
\label{equation4az}
\end{equation}

For convenience, we use the notation
$$
\text{Re} \,z=:x, \; \text{Im} \,z=: y, \; |z| =: r \;\;\text{and} \;\;
\arg z =: \theta
$$
and  set
\begin{equation}
u(x,y) := \frac{1}{|z|^4} f_\alpha(1/z)\,.
\label{u}
\end{equation}
By Equation (\ref{equation4fWhenAlphaIsNotZero}), we then have
\begin{equation}
u(x,y) = \int_0^{a(x,y)} f_\alpha 
\left (z-a \e^{\i \alpha} \right )\,\gamma_{p,s} (a)\,\d a
\label{equation4u}
\end{equation}
where 
$$
a(x,y) = 
\frac{x \sin \alpha+y \cos \alpha}{\sin(2 \alpha)}\,.
$$

For the sake of clarity, let us begin by considering
the case $p=1$.
In order to exploit the fact that $\gamma_{1,s}$ solves 
Equation  (\ref{ode4gamma}), we introduce the first-order
differential operator
\begin{equation}
\partial_\alpha := \cos \alpha \frac{\partial}{\partial x}
+ \sin \alpha \frac{\partial}{\partial y}
\label{D}
\end{equation}
and apply it to both sides of Equation (\ref{equation4u}).  Using
$$
\partial_\alpha a (x,y) = 1,
$$
we obtain
\begin{multline}
\notag
\partial_\alpha u(x,y) =
f_\alpha(z-a(x,y)\e^{\i \alpha})
\,\gamma_{1,s}(a(x,y)) + \int_0^{a(x,y)}
\partial_\alpha f_\alpha(z-a\,\e^{\i \alpha}) \gamma_{1,s}(a)
\,\d a \\
=
f_\alpha(z-a(x,y)\e^{\i \alpha})
\,\gamma_{1,s}(a(x,y)) - \int_0^{a(x,y)}
\frac{\d}{\d a}
f_\alpha(z-a\,\e^{\i \alpha}) \gamma_{1,s}(a)
\,\d a \\
=
f_\alpha(z)
\,\gamma_{1,s}(0) + \int_0^{a(x,y)}
f_\alpha(z-a\,\e^{\i \alpha})
\frac{\d \gamma_{1,s}}{\d a}(a)
\,\d a\,.
\end{multline}
Hence, we have
\begin{equation}
\partial_\alpha u(x,y) + \frac{1}{s} u(x,y) 
= \frac{1}{s r^4} u \left ( \frac{x}{x^2+y^2},\frac{-y}{x^2+y^2} \right )\,.
\label{generalisedComtet}
\end{equation}

Now, the characteristics of the first-order operator $\partial_\alpha$ are the
curves of equation
$$
z(t) =  z(0) + t \,\e^{\i \alpha}\,.
$$
Thus, the functions that are constant along the characteristics
depend only on $y \cos \alpha-x \sin \alpha$. 
This suggests the introduction of the new independent variables
\begin{equation}
X = -\frac{x \sin \alpha+y \cos \alpha}{x^2+y^2} \quad
\text{and} \quad Y = y \cos \alpha - x \sin \alpha\,.
\label{XandY}
\end{equation}
Set
\begin{equation}
U(X,Y) := u(x,y)\,.
\label{uandU}
\end{equation}
Then, Equation (\ref{generalisedComtet}) gives
\begin{equation}
\frac{\partial U}{\partial X}(X,Y)
=  \frac{r^4}{2 x y s} \left [ r^{-4} U(Y,X)- U(X,Y)\right ]\,.
\label{comtetXY}
\end{equation}
Evaluating ${\partial U}/{\partial X}$ at $(Y,X)$ instead of
$(X,Y)$,
we deduce that
\begin{multline}
\notag
\frac{\partial U}{\partial X}(Y,X) =
\frac{-1}{2 x y s} \left [ r^4 U(X,Y)-U(Y,X)\right ]  \\
= \frac{r^4}{2xys} \left [ r^{-4} U(Y,X)- U(X,Y) \right ]\,.
\end{multline}
Hence, we have the symmetry property
\begin{equation}
\frac{\partial U}{\partial X}  (Y,X) = \frac{\partial U}{\partial X} (X,Y)\,.
\label{comtetSymmetryProperty}
\end{equation}
This property can be used to eliminate the term $U(Y,X)$ from
Equation (\ref{comtetXY}) but,
although the resulting second-order
partial differential equation for $U$
is linear, the presence of coefficients
makes it awkward to solve.
We shall seek to guess the form of the solution
from Equation (\ref{comtetXY}) instead.

The first observation is that,
when $1/s = 0$, the solution is independent of $X$. Secondly,
we know that the solution enjoys the symmetry property
(\ref{comtetSymmetryProperty}). Hence, we are led to the ansatz
\begin{equation}
U(X,Y) = C\,\Phi'(Y)\,\exp \left \{ \frac{1}{s} \left [ \Phi(X)+\Phi(Y)\right ]
\right \},
\label{generalisedAnsatz}
\end{equation}
where $\Phi$ is 
some univariate function to be determined, and the
prime symbol denotes differentiation. Equation
(\ref{comtetXY}) then gives
\begin{equation}
\notag
\frac{2 x y}{r^4} \Phi'(Y)  =
\frac{1}{r^{4}}
-\frac{\Phi'(Y)}{\Phi'(X)} \,.
\end{equation}
Set
$$
\varphi = \frac{1}{\Phi'}\,.
$$
Then
$$
r^4 \varphi(X) = \varphi(Y)- 2xy
$$
or, in terms of $x$ and $y$,
$$
(x^2+y^2)^2
\varphi \left ( -\frac{x \sin \alpha + y \cos \alpha}{x^2+y^2}\right )
= \varphi (y\cos \alpha-x \sin \alpha) - 2 x y\,.
$$
Let $\varphi(t) = c t^2$. Then
\begin{multline}
\notag
c \left ( x^2 \sin^2 \alpha + 2 x y \sin \alpha \cos
\alpha + y^2 \cos^2 \alpha \right )  \\
=
c \left ( x^2 \sin^2 \alpha - 2 x y \sin \alpha \cos
\alpha + y^2 \cos^2 \alpha \right ) - 2 x y
\end{multline}
and so
$$
c = \frac{-1}{\sin(2 \alpha)}\,.
$$
Hence, we have found
$$
\varphi(t) = \frac{-t^2}{\sin(2 \alpha)},
\quad \Phi'(t) = -\frac{\sin (2 \alpha)}{t^2}
$$
and so
$$
\Phi (t) = \frac{\sin(2 \alpha)}{t}\,.
$$
Reporting this in Equation (\ref{generalisedAnsatz}) we obtain
$$
U(X,Y) = -C\,\frac{\sin(2 \alpha)}{Y^2}
\,\exp \left \{ \frac{\sin(2 \alpha)}{s} \left [ \frac{1}{X}
+ \frac{1}{Y} \right ] \right \}\,.
$$
Hence, in terms of $f$, $r$ and $\theta$, we have found
that, when $a$ is $\gamma_{p,s}$-distributed with $p=1$,
then
\begin{equation}
\notag
f_\alpha(z) = \frac{c_{1,s}(\alpha) \sin ( 2 |\alpha|)}{r^2 \sin^2 (\alpha+\theta)}
\exp \left \{ -\frac{\sin(2 \alpha)}{s} \left [
 \frac{1}{r \sin(\alpha-\theta)}
+ \frac{r}{\sin(\alpha+\theta)}
\right ] \right \},
\end{equation}
where $c_{1,s}(\alpha)>0$ is a normalisation constant.

For $p=2,3, \ldots$, we find by a similar
calculation that the equation for $u$ is
\begin{equation}
\left [ \partial_\alpha+\frac{1}{s} \right ]^p u(x,y) = \frac{1}{r^4 s^p}
u \left ( \frac{x}{x^2+y^2},\frac{-y}{x^2+y^2}\right )\,.
\label{equation4largeP}
\end{equation}
It would be tedious to analyse
this equation in detail.
It is however natural to conjecture that,
as in the case $\alpha=0$, $u$ consists
of an exponential term independent of $p$,
multiplied by some
algebraic part independent of $s$. Thus, we seek a solution of the form
\begin{equation}
u(x,y) = g_{p} (x,y) \,\exp
\left \{ \frac{\sin(2 \alpha)}{s} \left [ \frac{1}{X}+
\frac{1}{Y} \right ]
\right \}\,.
\label{ansatz4largeP}
\end{equation}
By setting 
$1/s = 0$ in Equation (\ref{equation4largeP}), we find
$$
\partial_\alpha^p g_{p} (x,y) = 0\,.
$$
The general solution of this equation is
$$
g_{p}(x,y)
= \varphi_0(Y) + r^2 X \varphi_1(Y) + \ldots + \left ( r^2 X \right )^{p-1}
\varphi_{p-1} (Y),
$$
for some univariate functions $\varphi_k$, $0 \le k <p$. By substituting
the resulting expression for $u$ in
Equation (\ref{equation4largeP}), we obtain
$$
\varphi_k( t) = \begin{cases}
C\,t^{-p-1} & \text{if $k=p-1$} \\
0 & \text{otherwise}
\end{cases}
\,.
$$

These results are summarised and stated in the most general form in Theorem \ref{generalisedLetacSeshadriThm}. 
The proof is a simple calculation carried out in Appendix \ref{generalisedLetacSeshadriAppendix}.

In Appendix \ref{momentAppendix}, we show how to express the moments
of the distribution $\nu$ in terms of products of modified Bessel functions. 
From these calculations, we deduce in particular that
\begin{equation}
c_{p,s}(\alpha) = 
\frac{1}{\left | 2 K_p \left ( 2 \e^{\i \alpha}/s
\right ) \right |^{2}}\,.
\label{normalisationConstant}
\end{equation}
Furthermore, the mean and the variance are given respectively by
\begin{equation}
{\mathbb E} \left ( Z \right ) := \int_{S_\alpha} z \,\nu (\d z) 
= \frac{K_{p-1} \left ( 2 \e^{-\i \alpha}/s \right )}{K_p \left ( 2 \e^{-\i \alpha}/s \right )} 
\label{mean}
\end{equation}
and
\begin{multline}
\text{Var} \left (Z\right ) := {\mathbb E} \left (\left |Z\right|^2 \right ) 
- \left |{\mathbb E} \left (Z\right )\right |^2 
=  \frac{s}{\sin(2 \alpha)} \text{Im} \left \{ \e^{\i \alpha} 
\frac{K_{p-1} (2 \e^{-\i \alpha}/s)}{K_p (2 \e^{-\i \alpha}/s)} \right \}\,.
\label{variance}
\end{multline}

\section{The case $\alpha = \pm \pi/2$}
\label{alphaIsPiOver2}

For $|\alpha| = \pi/2$, we introduce a
function $f_{\pm} : {\mathbb R} \rightarrow {\mathbb R}$
defined by
\begin{equation}
f_{\pm \pi/2}(z)  = c_{p,s}(\pm \pi/2)\,{f}_{\pm}
(y) \delta(x) \quad \text{for $z=x+\i y$ with $x \ge 0$
and $y \in {\mathbb R}$},
\label{fWhenAlphaIsPiOver2}
\end{equation}
where $c_{p,s}(\pm \pi/2)$ is the normalisation constant and $\delta$ is the Dirac delta.
In order to find an equation for $f_{\pm}$, we write the iteration
(\ref{forwardIteration}) in terms of $Y_n \in {\mathbb R}$, where
$$
Z_n = \i Y_n\,.
$$
This gives
\begin{equation}
Y_{n} = \frac{-1}{\pm a_n + Y_{n-1}}, \quad n \in {\mathbb N}\,.
\label{forwardIterationWhenAlphaIsPiOver2}
\end{equation}
It is then straightforward to obtain
\begin{equation}
f_{\pm}(y) = 
\frac{1}{y^2} \int_0^\infty {f}_{\pm} \left ( \mp a
- \frac{1}{y} \right )
\,\mu (\d a), \quad y \in {\mathbb R}\,.
\label{equation4fWhenAlphaIsPiOver2}
\end{equation}
Since
\begin{equation}
\forall \, y \in {\mathbb R}, \quad {f}_{-}(y) = {f}_{+}(-y),
\label{AlphaIsPiOver2}
\end{equation}
we need only consider the case $\alpha = \pi/2$.

As in the previous section, we shall seek to replace 
(\ref{equation4fWhenAlphaIsPiOver2}) by a differential equation. Set
$$
u(y) := \frac{1}{y^2} {f}_{+}(-1/y)\,.
$$
Then, Equation (\ref{equation4fWhenAlphaIsPiOver2}) gives
$$
u(y) = \int_0^\infty {f}_{+} (y-a)\,\gamma_{p,s}(a) \, \d a\,.
$$
Differentiating with respect to $y$, we obtain
\begin{multline}
\frac{\d u}{\d y} (y) = -\int_0^\infty \frac{\d}{\d a}
{f}_{+}(y-a) \gamma_{p,s}(a) \,\d a \\
= -{f}_{+}(y-a) \gamma_{p,s}(a) \left |_0^\infty \right .
+ \int_0^\infty {f}_{+}(y-a) \frac{\d \gamma_{p,s}}{\d a}(a)\,\d a \\
= {f}_{+}(y) \gamma_{p,s}(0)
+ \int_0^\infty 
{f}_{+}(y-a) \frac{\d \gamma_{p,s}}{\d a}(a)\,\d a\,. 
\notag
\end{multline}
By using the fact that $\gamma_{p,s}$ satisfies
(\ref{ode4gamma}) and repeating this calculation, 
we readily deduce the equation
\begin{equation}
\left [\frac{\d}{\d y} + \frac{1}{s} \right ]^p u(y)
= \frac{1}{s^p} \frac{1}{y^2} u(-1/y)\,.
\label{equation4uWhenAlphaIsPiOver2}
\end{equation}

Again, in order to solve this equation, it is useful to consider the
case $p=1$ first. We then have
\begin{equation}
y \frac{\d u}{\d y}(y) = \frac{1}{s} \left [ \frac{1}{y} u(-1/y)
-y u(y) \right ],
\label{anotherEquation4uWhenAlphaIsPiOver2}
\end{equation}
and so we deduce the property
\begin{equation}
- \frac{1}{y} \frac{\d u}{\d y} (-1/y) = y \frac{\d u}{\d y}(y)\,.
\label{symmetryWhenAlphaIsPiOver2}
\end{equation}
This property can be used to eliminate the $u(-1/y)$ term in
Equation (\ref{anotherEquation4uWhenAlphaIsPiOver2}). Multiplying the
equation by $y$, differentiating, and then multiplying by $-y$, we obtain
\begin{multline}
\notag
-y \frac{\d}{\d y} \left [ y^2 \frac{\d u}{\d y} (y) \right ]
= \frac{1}{s} \left \{- \frac{1}{y} \frac{\d u}{\d y} (-1/y) 
+ y \frac{\d}{\d y} \left [ y^2 u(y) \right ] \right \} \\
= \frac{1}{s} \left \{y  \frac{\d u}{\d y} (y) 
+ y \frac{\d}{\d y} \left [ y^2 u(y) \right ] \right \}\,.
\end{multline}

Hence, after simplification and integration, we find
$$
-y^2 \frac{\d u}{\d y} = \frac{1}{s} \left ( 1+y^2\right ) u + C
$$
for some constant $C$; integrating again, we obtain
$$
u(y) = C\, \exp \left [ \frac{1}{s} 
\left ( \frac{1}{y}-y \right ) \right ] \int^y t^{-2} 
\exp \left [ -\frac{1}{s} 
\left ( \frac{1}{t}-t \right ) \right ]\,\d t\,.
$$
In terms of ${f}_{+}$, this gives, after some simple manipulations,
\begin{equation}
\notag
{f}_{+}(y) =  \frac{1}{y^{2}} 
\exp \left [ \frac{1}{s} 
\left ( \frac{1}{y}-y \right ) \right ] \int_{c(y)}^y 
\exp \left [ -\frac{1}{s} 
\left ( \frac{1}{t}-t \right ) \right ]\,\d t
\end{equation}
where
 $$
 c(y)= \begin{cases}
-\infty & \text{if $y<0$} \\
0 & \text{if $y>0$}
\end{cases}\,.
$$

This result generalises to arbitrary positive values of $p$; see Theorem \ref{alphaIsPiOver2Theorem},
which is proved in Appendix \ref{P2Thm}.

\begin{remark}
The rigorous determination of the normalisation constant requires some
care, and we shall in fact make use of a weak limit result which will be
discussed in the next section. For the moment, let us merely point
out that the function $f_{\pm}$ thus defined is integrable.
Indeed, it is readily verified (by use of L'Hospital's rule) that the function
is continuous at $y=0$. 
By construction, it satisfies
Equation (\ref{equation4fWhenAlphaIsPiOver2}). Hence
$$
{f}_{\pm}(y) = O(y^{-2}) \quad \text{as $|y| \rightarrow
\infty$}\,.
$$
This shows that $f_{\pm} \in L^1 ({\mathbb R})$. We shall see in due course
that
\begin{equation}
c_{p,s}(\pi/2) = \lim_{\alpha \rightarrow \pi/2-} \frac{1}{4 \left | 
K_p \left ( 2 \e^{\i \alpha}/s \right ) \right |^2} 
= \frac{1}{\pi^2 \left [ J_p^2(2/s)+Y_p^2(2/s) \right ]}\,.
\label{normalisationConstantForPiOver2}
\end{equation}
\label{normalisabilityWhenAlphaIsPiOver2}
\end{remark}

Plots of $c_{p,s}(\pi/2) f_{+}$
for various choices of the parameters $s$ and $p$ are shown
in Figure \ref{piOver2PdfFigure}.
We can gain some insight
into the behaviour of the random variable $Z$ by considering
the limit $s \rightarrow 0$ (cf. Appendix \ref{lyapunovAppendix}). 
Set 
$$
\varphi (t) = t - \frac{1}{t}
$$
and note that
$$
1 = \frac{t^2}{t^2+1} \,\varphi'(t)\,.
$$
Then, using integration by parts,
\begin{multline}
\notag
\int_{c(y)}^y \exp \left [ -\frac{1}{s} 
\left ( \frac{1}{t}-t \right ) \right ]
t^{p-1}\,\d t
= s \int_{c(y)}^y \frac{1}{s} \varphi'(t) \exp
\left [\frac{1}{s} \varphi (t) \right ] \frac{t^{p+1}}{1+t^2}\,
\d t \\
= s \exp \left [ \frac{1}{s} \varphi (y) \right ] 
\frac{y^{p+1}}{1+y^2}
+ s \int_{c(y)}^y \exp \left [ \frac{1}{s} \varphi (t) \right ]
\frac{\d}{\d t} \left ( - \frac{t^{p+1}}{1+t^2} \right ) \d t\,. 
\end{multline}
Iterating, we readily obtain
\begin{equation}
f_{+}(y) \sim \sum_{n=1}^\infty  f_{+}^{(n)}(y) \,s^n
\quad \text{as $s \rightarrow 0$},
\label{asymptotics4PiOver2}
\end{equation}
where
\begin{equation}
f_{+}^{(1)}(y) = \frac{1}{1+y^2} \quad \text{and} \quad
f_{+}^{(n+1)}(y) = \frac{-1}{y^{p-1} (1+y^2)} \frac{\d}{\d y}
\left [ y^{p+1} f_{+}^{(n)}(y) \right ]\,.
\label{recurrence4fn}
\end{equation}
So by taking into account that (cf. \cite{Wa}, \S 13.74)
$$
\pi^2 \left [ J_p^2(2/s)+Y_p^2(2/s) \right ] 
\sim \pi s \quad \text{as $s \rightarrow 0$},
$$ 
we see that, for $s$ small, $Z$ is approximately Cauchy-distributed
along the imaginary axis; Figure \ref{piOver2PdfFigure} (a) provides a
graphical illustration of this fact.

\section{Weak limits of the random variable}
\label{weak}

Given the Letac--Seshadri result expressed in
Equation (\ref{letacSeshadriDistribution}),
the form of the density function for $\alpha \ne 0$, given
in Theorem \ref{generalisedLetacSeshadriThm}, may seem somewhat surprising.
In this section, we show that the random variable $Z$ has a weak limit
as $\alpha \rightarrow 0+$, and that the density function of
the limit is indeed given by the formula (\ref{letacSeshadriDistribution}).
In the same way, it will be shown that $Z$ has weak limits as 
$|\alpha| \rightarrow \pi/2-$ whose density functions are
those found in \S \ref{alphaIsPiOver2}. In fact, it was by
considering this weak limit that we were able to discover
the formula for ${f}_{\pm \pi/2}$ when $p>1$.

For this purpose, it will be convenient to extend the definition
of $f_\alpha$ to the whole of the complex plane as follows:
$$
\forall \, z \notin S_\alpha, \quad f_\alpha(z) = 0\,.
$$
Then, we define a
probability measure $\nu_\alpha$ on sets $S \subseteq {\mathbb C}$ via
$$
\nu_{\alpha} (S) =
\int_S  f_\alpha (z) \d z, \quad 0 < |\alpha| \le \pi/2\,.
$$
We also define the following singular probability measure
on ${\mathbb C}$: 
$$
\nu_0 (S) = \frac{1}{2 K_p (2/s)} \int_{S \cap {\mathbb R}_+}
x^{-p-1} 
\exp{\left [ -\frac{1}{s} \left ( x + \frac{1}{x} \right ) \right ] }
\d x\,.
$$
We shall use the notation $\xrightarrow[]{W}$ to indicate a
weak limit (limit in distribution).

\begin{theorem}
\label{weakLimitThm}
Let $X$, $\text{{\em i}} Y_{\pm}$ and $Z_\alpha$ be random variables with probability
measures $\nu_0$, $\nu_{\pm \pi/2}$ and $\nu_\alpha$
respectively. Then
$$
Z_\alpha \xrightarrow[\alpha \rightarrow 0]{W} X
\quad \text{and}
\quad Z_\alpha \xrightarrow[\alpha \rightarrow \pm \pi/2\mp]{W} \text{{\em i}} Y_{\pm}\,.
$$
\end{theorem}

\begin{proof}

Let $g : {\mathbb C} \rightarrow {\mathbb R}$ be an arbitrary
bounded and
continuous function. It will be necessary and sufficient 
(see \cite{Sh}) to show that
\begin{equation}
\int_{\mathbb C} g(z)  f_{\alpha}(z) \d z \xrightarrow[\alpha
\rightarrow \ell]{}
\int_{\mathbb C} g(z) \nu_{\ell} ( \d z ), \quad \ell = 0, \, 
\pm \pi/2\mp\,.
\label{weakLimit}
\end{equation}
Furthermore, since
$$
f_{-\alpha}(z) = f_{\alpha}(\overline{z}),
$$
we shall only consider positive values of $\alpha$.
 
We express the integral on the left-hand side in polar coordinates, and then
make the substitution
\begin{equation}
t=r\frac{\sin(\alpha - \theta)}{\sin(\alpha +\theta)} > 0\,.
\label{tSubstitution}
\end{equation}
Note that
\begin{equation}
\notag
\d t = -\frac{r\sin(2\alpha)}{ \sin^2(\alpha + \theta)} \d \theta,
\end{equation}
\begin{equation}
\notag
t+\frac{1}{t} + r+\frac{1}{r}=\frac{\cos\theta}{\cos\alpha}
\left[ r \frac{\sin(2 \alpha)}{ \sin(\alpha +\theta) }
+ \frac{1}{r} \frac{\sin(2 \alpha)}{ \sin(\alpha -\theta) }
 \right],
\end{equation}
\begin{equation}
\notag
\sin{\theta}= \frac{\text{sgn}(r-t)}{ \sqrt{ 1+ \left ( \frac{r+t}{r-t} \right ) ^2  
\cot^{2}\alpha }}
\xrightarrow[\alpha \rightarrow \ell]{} 
\begin{cases}
0 & \text{if $\ell=0+$} \\
\text{sgn}(r-t) & \text{if $\ell=\pi/2-$}
\end{cases}
\end{equation}
and
\begin{equation}
\notag
\cos{\theta}= \frac{1}{\sqrt{ 1+ \left ( \frac{r-t}{r+t} \right ) ^2  \tan^{2}\alpha }}
\xrightarrow[\alpha \rightarrow \ell]{} 
\begin{cases}
1 & \text{if $\ell =0+$} \\
0 & \text{if $\ell = \pi/2-$} 
\end{cases}\,.
\end{equation}
So, the substitution (\ref{tSubstitution}) yields
\begin{equation}
\int_{{\mathbb C}} g(z) f_{\alpha}(z) \d z
= c_{p,s}(\alpha)
\int_0^\infty \int_0^\infty \varphi_{\alpha}(r,t) \d t \d r,
\label{doubleIntegral}
\end{equation}
where
\begin{equation}
\notag
\varphi_{\alpha} := g \circ z(r,t) \,
\frac{t^{p-1}}{r^{p+1}}
\exp \left [ -\frac{1}{s}
\sqrt{ \cos^2 \alpha+ \left ( \frac{r-t}{r+t} \right ) ^2  \sin^{2}\alpha }
\left ( t+\frac{1}{t} + r+\frac{1}{r} \right ) \right ] 
\end{equation}
and 
$$
c_{p,s}(\alpha) = \frac{1}{\left | 2 K_p (2 \e^{\i \alpha}/s) \right |^2}\,.
$$

Our intention is to let $\alpha$ tend to its limit
and take the limit on the right-hand side of Equation (\ref{doubleIntegral})
under the integral sign. For this purpose, we need to find an integrable
function that provides an upper bound for the family $\{ \varphi_{\alpha} \}$.
By using the elementary inequality
$$
\sqrt{ \cos^2\alpha+ \left ( \frac{r-t}{r+t} \right ) ^2  \sin^{2}\alpha }
\ge \min{\left\{ 1, \frac{|r-t|}{r+t} \right\}} =\frac{|r-t|}{r+t},
$$
we obtain 
\begin{equation}
\notag
0 \le \varphi_{\alpha} \le \varphi,
\end{equation}
where $\varphi : {\mathbb R}_+^2 \rightarrow {\mathbb R}$ is defined by
$$
\varphi(r,t) = \frac{t^{p-1}}{r^{p+1}}  
\exp \left [ -\frac{1}{s} \frac{|r-t|}{r+t}
\left ( t+\frac{1}{t} + r+\frac{1}{r} \right ) \right ]
\| g \|_{L^\infty ({\mathbb C})} \,.
$$
Furthermore, as will become clear very shortly when we consider the limit
$\alpha \rightarrow \pi/2-$, we have
\begin{multline}
\notag
\| g \|_{L^\infty ({\mathbb C})}^{-1}
\int_0^\infty \int_0^\infty  \varphi (r,t) \d r \d t = \\
 \int_{\mathbb R} \frac{1}{y^{p+1}}
\exp \left [ \frac{1}{s} \left ( \frac{1}{y}-y \right ) \right ]
\int_{c(y)}^y \exp \left [ -\frac{1}{s}
\left ( \frac{1}{t}-t \right ) \right ] t^{p-1} \d t \d y \\
= c_{p,s}(\pi/2)^{-1}\,.
\end{multline}
We conclude (see Remark
\ref{normalisabilityWhenAlphaIsPiOver2}) that $\varphi$ is integrable and
so
we can apply Lebesgue's Dominated Convergence Theorem:
\begin{multline}
\lim_{\alpha \rightarrow 0+} \int_{\mathbb C} g(z) \nu_\alpha(\d z) 
= \lim_{\alpha \rightarrow 0+} c_{p,s}(\alpha) \,\lim_{\alpha \rightarrow 0+}
\int_{0}^\infty \int_0^{\infty} \varphi_{\alpha}(r,t) \d t \d r  \\ =
\lim_{\alpha \rightarrow 0+} c_{p,s}(\alpha)\,\int_{0}^\infty \int_0^{\infty} \lim_{\alpha \rightarrow 0+}
\varphi_{\alpha}(r,t) \d t \d r \\
= \frac{1}{4 K_p^2(2/s)} \int_0^\infty \int_0^{\infty}  g (r) \,
\frac{t^{p-1}}{r^{p+1}}  
\exp{\left ( -\frac{1}{s} \left[ t+\frac{1}{t}
+ r+\frac{1}{r} \right] \right )} \d t \d r\,.
\label{lebesgueLimit1}
\end{multline}
The double integral on the right-hand side gives
\begin{equation}
\notag
\int_{0}^\infty t^{p-1} \exp \left \{-\frac{1}{s} (t+1/t) \right \} \d t\,
\int_{0}^\infty r^{-p-1} \exp \left \{-\frac{1}{s} (r+1/r) \right \} g(r) \d r\,.
\end{equation}
By using the substitution $u=1/t$, and reporting the result in Equation (\ref{lebesgueLimit1}),
we obtain
\begin{multline}
\lim_{\alpha \rightarrow 0+} \int_{\mathbb C} g(z) \nu_\alpha(\d z) 
= \frac{1}{4 K_p^2(2/s)}
\left [ 2 K_p(2/s) \right ]^2 \int_{{\mathbb C}} g(z) \nu_0 (\d z ) \\
=\int_{{\mathbb C}} g(z) \nu_0 (\d z )\,.
\label{weakLimitAsAlphaTendsToZero}
\end{multline}

Similarly,
\begin{multline}
\lim_{\alpha \rightarrow \pi/2-} \int_{{\mathbb C}} g(z) \nu_\alpha (\d z) =
\lim_{\alpha \rightarrow \pi/2-} c_{p,s}(\alpha)\,\lim_{\alpha \rightarrow \pi/2-}
\int_{0}^\infty \int_0^{\infty} \varphi_{\alpha}(r,t) \,\d t \d r  \\
= \lim_{\alpha \rightarrow \pi/2-} c_{p,s}(\alpha)\, 
\int_{0}^\infty \int_0^{\infty} \lim_{\alpha \rightarrow \pi/2-} \varphi_{\alpha}(r,t) \,\d t \d r \\
= \frac{1}{\pi^2 \left [ J_p^2(2/s)+Y_p^2(2/s) \right ]} \\
\times
\int_{0}^\infty \int_0^ \infty g(\text{sgn}(r-t) r)
\frac{t^{p-1}}{r^{p+1}}
\exp \left \{ -\frac{1}{s} \frac{|r-t|}{r+t}
\left[ t+\frac{1}{t} + r+\frac{1}{r} \right] \right \}\, 
\d t \d r\,.
\label{lebesgueLimit2}
\end{multline}
The double integral on the right-hand side equals
\begin{multline}
\notag
\int_{0}^\infty \int_{0}^{r} g(r) \,\frac{t^{p-1}}{r^{p+1}}  
\exp{\left ( -\frac{1}{s} \left[ r- t + \frac{1}{t} - \frac{1}{r} \right] \right )}
\d t \d r  \\
+
\int_{0}^\infty \int_{r}^{\infty} g(-r) \,\frac{t^{p-1}}{r^{p+1}}  
\exp{\left ( -\frac{1}{s} \left[ t -r + \frac{1}{r} - \frac{1}{t} \right] \right )}
 \d t \d r \\
=
\int_{0}^{\infty} \int_{0}^{r} g(r) \,
t^{p-1}\exp{\left ( \frac{1}{s} \left[t - \frac{1}{t}\right]\right )} \d t 
  \frac{1}{r^{p+1}}  \exp{\left ( -\frac{1}{s}
\left[r - \frac{1}{r}\right]\right )}  \d r \\
+
\int_{-\infty}^{0}\int_{-\infty}^{r} g(r)\,
(-t)^{p-1}\exp{\left ( \frac{1}{s} \left[t - \frac{1}{t}\right]\right )} \d t
  \frac{1}{(-r)^{p+1}}  \exp{\left ( -\frac{1}{s}
  \left[r - \frac{1}{r}\right] \right )} dr\\
= \int_{{\mathbb R}} g(y)\,
   \frac{1}{y^{p+1}} \exp{ \left \{ -\frac{1}{s} \left[ y - \frac{1}{y} \right]
   \right \}}
\int_{c(y)}^y t^{p-1} \exp{ \left ( \frac{1}{s}
\left[ t - \frac{1}{t} \right] \right )} \d t \d y \\
= c_{p,s} (\pi/2)^{-1}
\,\int_{{\mathbb C}}
g(z) \,\nu_{\pi/2} (\d z)\,.
\end{multline}
Hence, after reporting this in Equation (\ref{lebesgueLimit2}), we obtain
\begin{equation}
\notag
\lim_{\alpha \rightarrow \pi/2-} \int_{\mathbb C} g(z) \nu_\alpha (\d z)
= \frac{c_{p,s} (\pi/2)^{-1}}{\pi^2 \left [ J_p^2(2/s)+Y_p^2(2/s) \right ]}
\int_{\mathbb C} g(z) \nu_{\pi/2} (\d z)\,.
\end{equation}  
By setting $g \equiv 1$ in that equation, we deduce that
$$
c_{p,s} (\pi/2) = \frac{1}{\pi^2 \left [ J_p^2(2/s)+Y_p^2(2/s) \right ]}
$$
and the proof is complete.
\end{proof}

\section{Calculation of the Lyapunov exponent}
\label{lyapunov}

The Lyapunov exponent
associated with the random continued fraction (\ref{continuedFraction})
features prominently in the applications considered in the introduction.
By using the formulae given
in Theorems \ref{generalisedLetacSeshadriThm}
and \ref{alphaIsPiOver2Theorem}, its computation 
reduces to a problem of quadrature. 
In this section, we prove that the Lyapunov exponent 
$\lambda_{p,s}(\alpha)$ is in fact given by the logarithmic derivative of an appropriate
modified Bessel function for arbitrary $p,s>0$ and $\alpha\in [-\pi/2,\pi/2]$.
In particular, the result yields explicit formulae for $\lambda_{p,s}(\alpha)$ 
when $p \in {\mathbb N}$, and simple asymptotic expansions for small and large $s$ 
when $p > 0$.

\begin{theorem}
\label{LyEx}
For any $p,s>0$ and $\alpha\in [-\pi/2,\pi/2]$, the Lyapunov exponent $\lambda_{p,s}(\alpha)$ 
associated with the invariant measure $\nu$ takes the form
\begin{equation}
\label{LyExp}
\lambda_{p,s}(\alpha)=\Real 
\frac{\partial_{p} K_{p}\left(\frac{2}{s} \text{\em e}^{\text{\em i}\alpha}\right)}{K_{p}\left(\frac{2}{s} \text{\em e}^{\text{\em i}\alpha}\right)}.
\end{equation}
\end{theorem}

\begin{proof}
Using equation (\ref{lyapunovIntegral}) and the substitution
$$
t = \frac{\sin (\alpha-\theta)}{\sin ( \alpha+\theta)}, \quad \rho = r \sqrt{t}, \quad  \varphi(t) = t + 1/t + 2 \cos(2\alpha),
$$
we have
\begin{multline}
\left | 2 K_p \left ( 2 \e^{\i \alpha }/s \right ) \right |^2 \lambda_{p,s}(\alpha) 
=-\left| 2 K_p \left ( 2 \e^{\i \alpha }/s \right ) \right |^2 \int_{S_{\alpha}}\ln|z|f_{\alpha}(z)dz \\
= \int_0^\infty \int_0^\infty \left [ \frac{1}{2} \ln t - \ln \rho \right ] t^{p-1}
\exp \left \{ -\frac{\sqrt{\varphi (t)}}{s} (\rho+1/\rho )\right \} \frac{\d \rho}{\rho} \d t \\
= \int_0^\infty \ln t \,t^{p-1} \frac{1}{2} \int_0^\infty \exp \left \{ 
-\frac{\sqrt{\varphi (t)}}{s} (\rho+1/\rho )\right \}  \frac{\d \rho}{\rho} \d t\,.
\notag
\end{multline}
Now using the relation
$$
\ln t \, t^{p-1}=\frac{\partial}{\partial p} t^{p-1} 
$$
together with the uniform convergence (for $p$ in compact sets) 
of the above integral and reversing the change of variables we can continue as follows

\begin{multline}
\left | 2 K_p \left ( 2 \e^{\i \alpha }/s \right ) \right |^2 \lambda_{p,s}(\alpha) 
= \frac{1}{2} \frac{\partial}{\partial p} \int_0^\infty  \,t^{p-1}  \int_0^\infty \exp \left \{ 
-\frac{\sqrt{\varphi (t)}}{s} (\rho+1/\rho )\right \}  \frac{\d \rho}{\rho} \d t \\
=\frac{1}{2} \frac{\partial}{\partial p} \left | 2 K_p \left ( 2 \e^{\i \alpha }/s \right ) \right |^2
=4\text{Re} \left( \frac{\partial}{\partial p} K_p \left ( 2 \e^{\i \alpha }/s \right )
K_p \left ( 2 \e^{-\i \alpha }/s \right ) \right).
\notag
\end{multline}

Hence,
\begin{multline}
\lambda_{p,s}(\alpha)=\frac{\Real \left( \frac{\partial}{\partial p} K_p \left ( 2 \e^{\i \alpha }/s \right )
K_p \left ( 2 \e^{-\i \alpha }/s \right ) \right)}{\left | K_p \left ( 2 \e^{\i \alpha }/s \right ) \right |^2}
 = \\  \Real \frac{\frac{\partial}{\partial p} K_p \left ( 2 \e^{\i \alpha }/s \right )}
{K_p \left ( 2 \e^{\i \alpha }/s \right )} =
\Real \frac{\partial}{\partial p} \ln K_{p}\left(\frac{2}{s} \e^{\i\alpha}\right)\,.
\notag
\end{multline}
\end{proof}

In the remarks below, we mention some of the most important consequences of the above theorem.

\begin{remark}
It is easy to derive a general recurrence relation for the Lyapunov exponent.
To this end, it is useful to define its complex generalisation as follows:
\begin{equation}
\label{CLyExn}
\Lambda_{p}(w):=   {\partial_{p} } \ln K_{p}(w)  = \frac{\partial_{p}K_{p}(w)}{K_{p}(w)}\,.
\end{equation}
Differentiating with respect to $p$ the following well-known recurrence relation 
for the modified Bessel function (see \cite{Wa} \S 3.71)
$$
K_{p}(w)=\frac{2(p-1)}{w}K_{p-1}(w)+K_{p-2}(w),
$$
one obtains the recurrence relation for $\Lambda_{p}(w)$
\begin{equation}
\Lambda_{p}(w)=
\frac{2(p-1)}{w}\frac{K_{p-1}(w)}{K_{p}(w)}
\Lambda_{p-1}(w)+ 
\frac{K_{p-2}(w)}{K_{p}(w)} \Lambda_{p-2}(w) +
\frac{2}{w}\frac{K_{p-1}(w)}{K_{p}(w)}
\end{equation}
which, in view of the fact that 
$$
\lambda_{p,s}(\alpha)= \Real \Lambda_{p}\left(\frac{2}{s}e^{i\alpha}\right),
$$
provides an efficient means of computing $\lambda_{p,s}(\alpha)$.
\end{remark}

\begin{figure}[htbp]
\vspace{8.5cm} 
\includegraphics{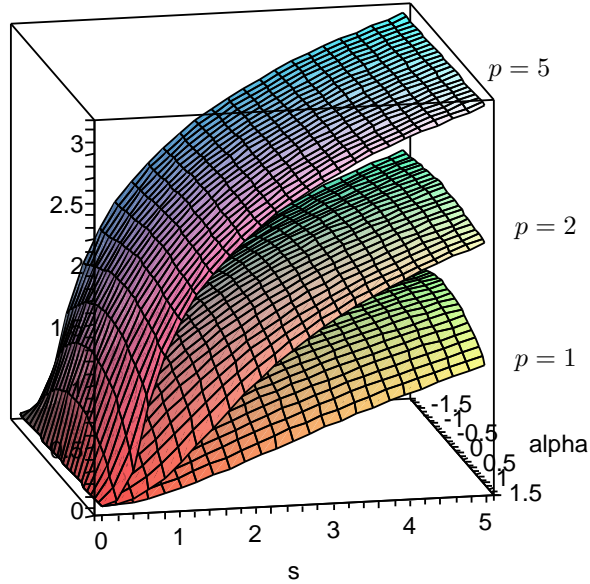}  
\begin{picture}(0,0) 
\put(90,80){$p=1$}
\put(90,130){$p=2$}
\put(80,190){$p=5$}
\end{picture} 
\caption{The Lyapunov exponent
$\lambda_{p,s}(\alpha)$ as a function of $s$ and $\alpha$ for $p=1,\,2,\,5$. }
\label{LyExFigure} 
\end{figure}

\begin{remark}
When $p=n\in \mathbb{N}$, we have
\begin{equation}
\label{LyExn}
\lambda_{n,s}(\alpha)= 
  \sum_{k=0}^{n-1} \frac{n!}{2(n-k)k!} \left(\frac{\e^{\i\alpha}}{s}\right)^{k-n}
 \frac{ K_{k}\left( \frac{2}{s} \e^{\i\alpha} \right)}
{K_{n}\left(\frac{2}{s} \e^{\i\alpha}\right) }\,.
\end{equation}
Indeed, the result follows from Theorem \ref{LyEx} and the formula (see \S 9.6.45 in \cite{AS})
$$
\left[ \frac{\partial}{\partial p} K_p(w) \right]_{p=n}= \sum_{k=0}^{n-1} \frac{n!}{2(n-k)k!} (w/2)^{k-n}K_{k}(w)\,.
$$

The two most important special cases of (\ref{LyExn}) are:
\begin{enumerate}
\item {\bf The Lyapunov exponent for the generalized inverse Gaussian law}.
\begin{equation}
\label{GIGLyap}
\lambda_{n,s}(0)= 
  \sum_{k=0}^{n-1} \frac{s^{n-k} n!}{2(n-k)k!}
\frac{K_{k}(2/s)}{K_{n}(2/s)}.
\end{equation}

\item {\bf The Lyapunov exponent for the Schr\"odinger case.}
Using the formula (see \cite{Wa} \S \S 3.6, 3.7)
$$
K_{n}(\i w)=\frac{\pi}{2} \e^{(3n+1)\frac{\pi \i}{2}} (J_{n}(w)+\i Y_{n}(w))
$$
one immediately obtains
\begin{equation}
\lambda_{n,s}(\pi/2)= 
  \sum_{k=0}^{n-1} \frac{s^{n-k}n!}{2(n-k)k!}
 \frac{ J_{k}(2/s) J_{n}(2/s) + Y_{k}(2/s) Y_{n}(2/s)}
{ J_{n}^{2}(2/s) + Y_{n}^{2}(2/s)} .
\end{equation}
\end{enumerate}
\end{remark}

Next, we investigate the asymptotic behaviour
of $\lambda_{p,s}(\alpha)$ for arbitrary positive $p$. In what follows, $\psi$ will denote the
{\em digamma function}.
We start with the large $s$ asymptotics. 

\begin{proposition}

For arbitrary $p>0$ and $\alpha\in [-\pi/2,\pi/2]$ we have the following large $s$ asymptotics
of the Lyapunov exponent
\begin{equation}
\label{LS}
\lambda_{p,s}(\alpha)
= \ln s + \psi(p) + R_{p,s}(\alpha),
\end{equation}
where
\begin{equation}
R_{p,s}(\alpha) =
\begin{cases}
  2\cos(2p\alpha)\frac{\Gamma(1-p)}{\Gamma(1+p)} \frac{\ln s}{s^2} + O \left ( 1/s^2 \right ), & \text{if $p \in (0,1)$} \\
   & \\
  2\cos (2\alpha) \frac{(\ln s)^2 }{s^2} + O \left ( \ln s / s^2 \right ), & \text{if $p=1$} \\
  & \\
  \frac{\cos (2\alpha)}{(p-1)^2} \frac{1}{s^2} + O \left ( E(s) \right ), &  \text{if $p>1$} \\
\end{cases}
\end{equation}
and
$$
E(s)= \begin{cases}
\ln s/s^{2p} & \text{if $p\in (1,2)$} \\
(\ln s)^2/s^{4} & \text{if $p=2$} \\
1/s^4 & \text{if $p>2$}
\end{cases}\,.
$$
\end{proposition}

\begin{remark}
Although only the leading term of the expansion for the remainder
$R_{p,s}$ is reported here,
it will be clear from the proof that Formula (\ref{LargeS}) 
allows one to compute arbitrarily many terms.
\end{remark}

\begin{proof} 
First, we note that, when $p \in {\mathbb N}$, the Lyapunov exponent is given
explicitly by Equation (\ref{LyExn}), and its asymptotics can be deduced directly from
the asymptotic expansion of the modified Bessel function of integral order
(see \S 8.446 in \cite{GR}).

So, from now on, we shall assume that $p$ is not an integer.
As a starting point, we take the following representations (Equations (8.485) and (8.445) in \cite{GR})
\begin{eqnarray}
\label{BesselK}
K_{p}(w)&=&\frac{\pi}{2}\frac{I_{-p}(w)-I_{p}(w)}{\sin(p \pi)}, \\
\label{BesselI}
I_{p}(w)&=& \left(\frac{w}{2} \right)^p \sum_{k=0}^{\infty} A_{k}(p) \left(\frac{w}{2} \right)^{2k},
\end{eqnarray}
where
$$
A_{k,p}:=\frac{1}{k!\Gamma(p+k+1)}
$$
These representations yield (cf. Equations (9.6.42-3) in \cite{AS})
\begin{eqnarray*}
\frac{\partial_{p}K_{p}(w)}{K_{p}(w)}&=&-\pi\cot(p\pi)+\frac{\partial_{p}I_{-p}(w)-\partial_{p}I_{p}(w)}{I_{-p}(z)-I_{p}(w)},\\
\partial_{p}I_{-p}(w)&=&-\ln\left(\frac{w}{2} \right)I_{-p}(w)+\left(\frac{w}{2} \right)^{-p} \sum_{k=0}^{\infty} B_{k}(-p) \left(\frac{w}{2} \right)^{2k},\\
\partial_{p}I_{p}(w)&=&\ln\left(\frac{w}{2} \right)I_{p}(w)-\left(\frac{w}{2} \right)^p \sum_{k=0}^{\infty} B_{k}(p) \left(\frac{w}{2} \right)^{2k},\\
\end{eqnarray*}
where
$$
B_{k,p}:=\frac{\psi(p+k+1)}{k!\Gamma(p+k+1)}\,.
$$
Thus
\begin{multline}
\Lambda_{p}(w) = -\pi\cot(p\pi) + \\
\frac{ \sum_{k=0}^{\infty}  \left(-A_{k,-p} \ln\left(\frac{w}{2} \right) +  B_{k,-p}\right)\left(\frac{w}{2} \right)^{2k}
  - \left( A_{k,p} \ln\left(\frac{w}{2} \right) - B_{k,p}  \right) \left(\frac{w}{2} \right)^{2(k+p)} }
{\sum_{k=0}^{\infty} A_{k,-p}  \left(\frac{w}{2} \right)^{2k} - A_{k,p}  \left(\frac{w}{2} \right)^{2(k+p)}}\,.
\notag
\end{multline}
For small $w$, we then have the asymptotic expansion
\begin{multline}
\label{LargeS}
\Lambda_{p}(w) \sim -\pi\cot(p\pi) + \Gamma(1-p)\times\\
 \left[ \sum_{k=0}^{\infty}  \left(A_{k,-p} \ln\left(\frac{2}{w} \right) +  
 B_{k,-p}\right)\left(\frac{w}{2} \right)^{2k} + \left( A_{k,p} \ln\left(\frac{2}{w} \right) 
 + B_{k,p}  \right) \left(\frac{w}{2} \right)^{2(k+p)}\right] 
 \times \\
 \left[ 1+ \sum_{n=1}^{\infty} \left( \sum_{k=0}^{\infty} \Gamma(1-p)A_{k,p}  \left(\frac{w}{2} \right)^{2(k+p)} - 
 \sum_{k=1}^{\infty} \Gamma(1-p)A_{k,-p}  \left(\frac{w}{2} \right)^{2k} \right)^n \right]\,.
\end{multline}

Applying the identities (cf. Equation (8.365), Points (1) and (8), \cite{GR})
$$ 
\psi(2-p)-\psi(1-p)=(1-p)^{-1}, \qquad \psi(1-p)-\pi\cot(p\pi) = \psi(p),
$$ 
and keeping only the most significant terms we obtain, for all
positive noninteger $p$ and small $w$, the expansion
\begin{multline}
\notag
\Lambda_{p}(w) \sim \ln(2/w) + \psi(p) + 
2\frac{\Gamma(1-p)}{\Gamma(1+p)}\ln(2/w)(w/2)^{2p}+
\frac{1}{(1-p)^2}(w/2)^{2}+ \cdots
\end{multline}
The behavior of the error term depends on the value of $p$. For noninteger $p$ we have
the following estimates of the error:  
$O(w^{2p})$ for $p\in(0,1)$; $O((\ln s)/w^{2p})$ for $p\in(1,2)$ and
$O(w^{4})$ for $p>2$.

Substituting $w=2 \e^{\i\alpha}/s$ and taking the real part (cf. Theorem \ref{LyEx}) completes the proof.
\end{proof}

It is worth noting that, when $p$ is large, there is an alternative way of
obtaining the asymptotics of the Lyapunov exponent. Indeed, by using the integral representation 
$$
K_{p}(z)=\frac{1}{2}\left( \frac{2}{z}\right)^p \int_{0}^{\infty}t^{p-1}e^{-t -\frac{z^2}{4t}} \,\d t,
$$
one easily deduces the following version of Equation (8.446) in \cite{GR}:
\begin{equation}
\label{KBessA}
K_{p}(z)\sim \frac{1}{2}\left( \frac{2}{z}\right)^p \sum_{0\leq k <p} \frac{\Gamma(p-k)}{k!}\left(- \frac{z^2}{4} \right)^k,
\quad  \text{as $z \rightarrow 0$}\,.
\end{equation}
This yields (\ref{LS}) for $p > 2$ and explains the absence of ``logarithmic'' prefactors in the 
low order terms of
the expansion for the Lyapunov exponent when $p$ is large. 

Next, we turn to the case of small $s$.

\begin{figure}[htbp]
\vspace{7.5cm} 
\begin{picture}(0,0) 
\put(-50,62){$p=1$}
\put(-80,92){$p=2$}
\put(-80,143){$p=3$}
\put(-85,178){$p=4$}
\put(-100,0){(a)}
\put(-70,10){$s$}
\put(140,48){$p=1/2$}
\put(140,83){$p=3/2$}
\put(145,143){$p=5/2$}
\put(145,182){$p=7/2$}
\put(100,0){(b)}
\put(130,10){$s$} 
\end{picture} 
\includegraphics{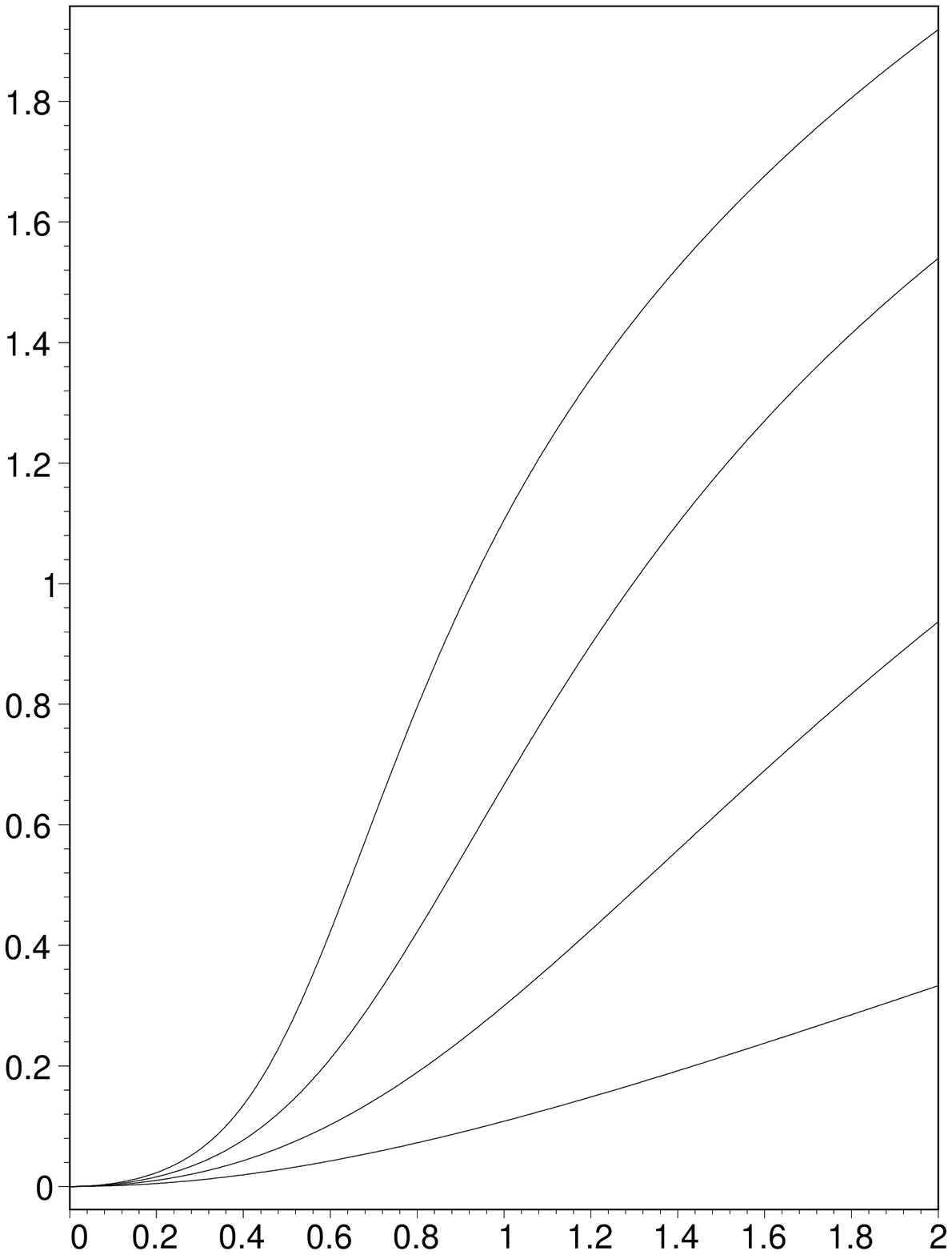}  
\includegraphics{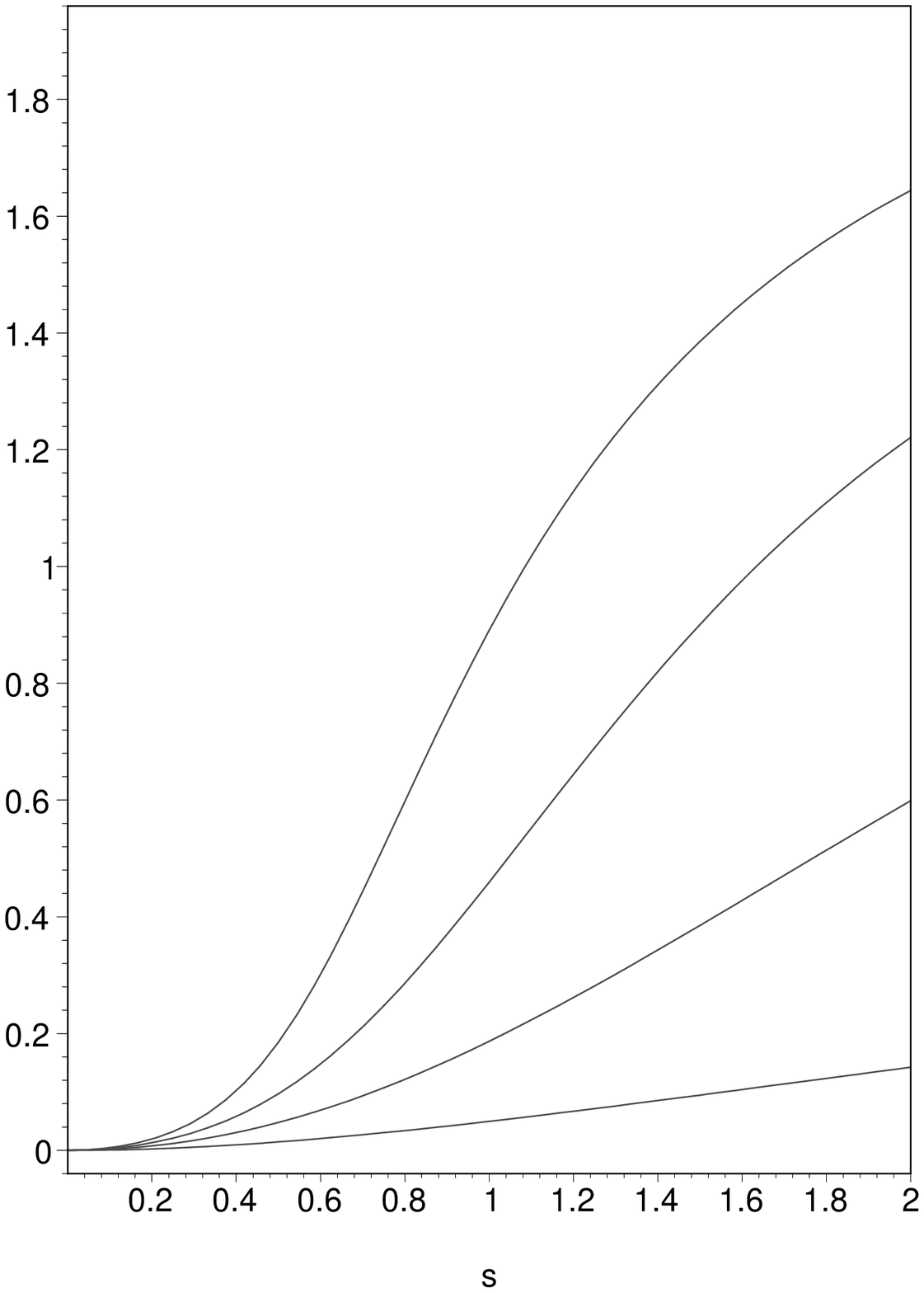} 
\caption{The Lyapunov exponent
$\lambda_{p,s}(\pi/2)$ as a function of $s$. The curves in (a) are
obtained by using the exact formula; those in (b) are obtained by summing the asymptotic
series as $s \rightarrow 0$.}
\label{piOver2LyapunovFigure} 
\end{figure}

\begin{proposition}

For arbitrary $p>0$ and $\alpha\in [-\pi/2,\pi/2]$, we have the following small $s$ asymptotics
of the Lyapunov exponent:
\begin{equation}
\label{lnK}
\lambda_{p,s}(\alpha) \sim \sum_{n=1}^{\infty}l_{n}s^{n} \quad \text{as $s \rightarrow 0$},
\end{equation}
where
\begin{multline}
l_{1}=\frac{p \cos \alpha}{2}, \quad l_{2}= - \frac{p  \cos 2\alpha}{8}, \quad  l_{3}=  - \frac{p(4p^2-13) \cos 3\alpha }{192}, \\
l_{4}= \frac{p(4p^2-7) \cos 4\alpha }{128}, \quad  l_{5} = \frac{p(-920p^2+48p^4+1187) \cos 5\alpha }{20480}\,.
\label{Coef}
\end{multline}
\end{proposition}
\begin{remark}
Higher order terms can be derived using formula (\ref{SmallS}) below.
\end{remark}
\begin{remark}
When $\alpha=\pm \pi/2$ (and only in this case) the real part of all odd order terms in (\ref{lnK}) vanish. 
Thus the leading order term in this particular case is quadratic (cf. Figures \ref{LyExFigure} and \ref{piOver2LyapunovFigure}).
\end{remark}

\begin{proof}
For small $s$, the asymptotics follow from the behaviour of the modified
Bessel functions for large $w$. We have (see \cite{Wa} \S 7.23)
$$
K_{p}(w) = \left( \frac{\pi}{2w}\right)^{1/2} \e^{-w} \sum_{m=0}^{n-1} \frac{(p,m)}{(2w)^{m}} + \Theta(p) \frac{(p,n)}{(2w)^{n}},
$$
where
$$
(p,m):=\frac{\Gamma(p+m+1/2)}{m!\Gamma(p-m+1/2)}=\frac{(4p^2-1)(4p^2-3^2)...(4p^2-(2m-1)^2)}{2^{2m}m!}
$$
and where the constant $\Theta$ satisfies 
$$
|\Theta|<1 \quad \text{and} \quad \text{Re} \Theta \geq 0
\qquad
\text{for $p \in{\mathbb R}$ and $\text{Re} w \geq 0$}\,.
$$
(If $n>p-1/2$, then $0\leq \Theta \leq 1$, see \cite{Wa} \S 7.30).
Thus, we have
\begin{equation}
\frac{\partial_{p}K_{p}(w)}{K_{p}(w)}= \frac{\sum_{m=0}^{n-1} \frac{\partial_{p}(p,m)}{(2w)^{m}} 
+  \frac{\partial_{p}(\Theta(p) (p,n))}{(2w)^{n}}}
{\sum_{m=0}^{n-1} \frac{(p,m)}{(2w)^{m}} + \Theta(p) \frac{(p,n)}{(2w)^{n}}}\,,
\end{equation}
where
$$
\partial_{p}(p,m)=\frac{\Gamma(p+m+1/2)}{m!\Gamma(p-m+1/2)}[\psi(p+m+1/2)-\psi(p+m+1/2)].
$$
This leads to the following expansion for large $w$
\begin{multline}
\label{SmallS}
\Lambda_{p}(w) \sim \left[\sum_{m=1}^{n-1} \frac{\partial_{p}(p,m)}{(2w)^{m}} +\frac{\partial_{p}(\Theta(p) (p,n))}{(2w)^{n}}\right]
\times\\ 
\left[1+ \sum_{N=1}^{\infty}\left(-\sum_{m=1}^{n-1} \frac{(p,m)}{(2w)^{m}}-  \Theta(p) \frac{(p,n)}{(2w)^{n}}\right)^N \right]\,.
\end{multline}
From this expansion, every coefficient in (\ref{lnK}) can be computed explicitly.
In particular, by using the basic properties of the gamma and digamma functions, we obtain 
\begin{multline}
\Lambda_{p}(w) \sim \frac{p}{w} -  \frac{p}{2w^2} - \frac{p(4p^2-13)}{24w^3} +\\ 
\frac{p(4p^2-7)}{8w^4} + \frac{p(-920p^2+48p^4+1187) }{640w^5}- \cdots \quad  \text{as $w \rightarrow \infty$}\,.
\end{multline}
By setting $w=2 \e^{\i\alpha}/s$ and taking the real part,
one obtains Equation (\ref{Coef}).
\end{proof}

\begin{remark}
Although the series (\ref{lnK}) is divergent, one can compute its sum by using standard 
summation techniques \cite{BO}. The plots of $\lambda_{p,s}(\pi/2)$ against $s$ for various
half-integer values of $p$ shown in Figure \ref{piOver2LyapunovFigure} (b)
were obtained by using a Pad\'{e} approximant--- i.e.
a rational function of $s$ that matches the series to $O(s^{64})$.
\end{remark}

\medskip
{\bf Acknowledgments:} We thank Brian Winn for a very helpful discussion leading to the discovery of
Formula (\ref{LyExp}).

\appendix

\section{Some identities involving products of Bessel functions}
\label{besselAppendix}

We begin with two integral representations of the modified Bessel function:
For $\Real z >0$, we have (cf. \cite{Wa}, \S 6.22) 
\begin{multline}
K_p (z) 
= \frac{1}{2} \int_0^\infty t^{-p-1} \exp \left \{ -\frac{z}{2} (t+1/t)
\right \} \, \d t \\
= \frac{z^p}{2} \int_0^\infty \tau^{-p-1} \exp \left \{ -\frac{1}{2} \left ( \tau +z^2/\tau \right ) \right \}\,\d \tau\,. 
\label{integralRepresentation}
\end{multline}

When calculating the moments of $\nu$, we will use integrals of the form 
\begin{equation}
{\mathcal I}_p^{(n)}(u,v) :=
\frac{1}{2} \int_0^\infty \exp \left \{ -\frac{\tau}{2}-\frac{u^2+v^2}{2 \tau} \right \} K_p \left ( \frac{uv}{\tau} \right ) 
\tau^n \,\d \tau, 
\label{macdonaldIntegral}
\end{equation}
where $n \in {\mathbb Z}$, and the complex numbers $u$ and $v$ are such that 
$$
| \arg u | < \pi, \quad |\arg v| < \pi \quad \text{and} \quad |\arg(u+v)| < \pi/4\,.
$$
For $n=-1$, the value is given by Macdonald's formula \cite{Wa}, \S 13.71:
\begin{equation}
{\mathcal I}_p^{(-1)}(u,v) = K_p (u) K_p (v)\,.
\label{macdonaldFormula}
\end{equation}
Differentiate this identity with respect to $u$, and then multiply it by $u$ to obtain
\begin{multline}
u K_p'(u) K_p(v) 
= \frac{1}{2} \int_0^\infty \exp \left \{ -\frac{\tau}{2} - \frac{u^2+v^2}{2 \tau} \right \}
\left [ - \frac{u^2}{\tau^2} K_p \left ( \frac{uv}{\tau} \right ) + \frac{uv}{\tau^2} 
K_p' \left ( \frac{uv}{\tau} \right ) \right ]\,\d \tau \\
= \frac{1}{2} \int_0^\infty \exp \left \{ -\frac{\tau}{2} - \frac{u^2+v^2}{2 \tau} \right \}
 \left [ -\frac{1}{2} + \frac{v^2-u^2}{2 \tau^2} \right ]  K_p \left ( \frac{uv}{\tau} \right ) \d \tau
\notag
\end{multline}
after integration by parts. If we interchange $u$ and $v$, this becomes
\begin{multline}
v K_p(u) K_p'(v) \\
= \frac{1}{2} \int_0^\infty \exp \left \{ -\frac{\tau}{2} - \frac{u^2+v^2}{2 \tau} \right \}
 \left [ -\frac{1}{2} - \frac{v^2-u^2}{2 \tau^2} \right ]  K_p \left ( \frac{uv}{\tau} \right ) \d \tau \,.
\notag
\end{multline}
Hence, by adding these two identities, we find 
\begin{equation}
{\mathcal I}_p^{(0)}(u,v)
= u K_{p+1}(u) K_p(v) + v K_{p+1}(v) K_p(u) -2 p \,K_p(u) K_p(v) \,.
\label{crossProduct}
\end{equation}
By similar calculations, we  obtain easily
the recurrence relations
\begin{equation}
{\mathcal I}_{p}^{(n)} = \left [ 2 n  - u \frac{\partial}{\partial u}
- v \frac{\partial}{\partial v} \right ]\,{\mathcal I}_{p}^{(n-1)},
\quad n \in {\mathbb N}\,.
\label{McdonaldForwardRecurrence}
\end{equation}
and
\begin{equation}
(v^2-u^2) {\mathcal I}_{p}^{(n-1)} = \left [ u \frac{\partial}{\partial u}
- v \frac{\partial}{\partial v} \right ]\,{\mathcal I}_{p}^{(n)},
\quad n \in {\mathbb Z}\,.
\label{McdonaldBackwardRecurrence}
\end{equation}
The first of these enables the evaluation of the integral (\ref{macdonaldIntegral})
in terms of products of modified Bessel functions when $n=0,1, \ldots$, whereas the second
is appropriate when $n = -2,-3,\ldots$ (provided that $u^2 \ne v^2$).

\section{Calculation of the moments}
\label{momentAppendix}
In this section, we show how to calculate the moments 
\begin{equation}
M_{p,s}^{(m,n)}(\alpha) := \int_{S_\alpha} z^m \bar{z}^n \,f(z) \,\d z, \quad m,n \in {\mathbb N},
\label{moments}
\end{equation}
of the distribution $\nu$. For simplicity, we shall assume that $\alpha > 0$.

We have
\begin{multline}
\notag
M_{p,s}^{(m,n)}(\alpha) = \int_{-\alpha}^\alpha
\int_0^\infty  r^{m+n} \e^{\i (m-n) \theta}\,\frac{c_{p,s}(\alpha) \sin(2 \alpha)}{r^2 \sin^2(\alpha+\theta)}
\left [ \frac{\sin(\alpha-\theta)}{\sin(\alpha+\theta)} \right ]^{p-1} \\
\times \exp \left \{ -\frac{\sin(2\alpha)}{s} \left [
\frac{1}{r \sin(\alpha-\theta)}+\frac{r}{\sin(\alpha+\theta)} \right ]
\right \}\, r \d r \d \theta \,.
\end{multline}
The substitution
$$
t = \frac{\sin(\alpha-\theta)}{\sin(\alpha+\theta)}
$$
yields
$$
\d t = - \frac{\sin(2 \alpha)}{\sin^2 (\alpha+\theta)}\,\d \theta,\;\;
\e^{\i \theta} = \frac{\e^{\i \alpha} + t \e^{-\i \alpha}}{\sqrt{t \varphi(t)}} 
$$
and
$$
\frac{2 \sin(2 \alpha)}{s \sqrt{\sin(\alpha-\theta) \sin(\alpha+\theta)}}
= \frac{2}{s} \sqrt{\varphi (t)},
$$
where
$$
\varphi (t) = t + 1/t + 2 \cos(2 \alpha)\,.
$$
Hence 
\begin{multline}
\notag
M_{p,s}^{(m,n)}(\alpha)=c_{p,s}(\alpha)
\int_{0}^\infty \int_0^\infty r^{m+n-1}t^{p-1} \left [ \frac{\e^{\i \alpha} + t \e^{-\i \alpha}}{\sqrt{t \varphi(t)}} \right ]^{m-n} \\
\times\exp\left\{-\frac{1}{s} \left [\frac{1}{r} \sqrt{\frac{ \varphi(t)}{t}} + r \sqrt{t \varphi(t)}\right ]\right\} \d r \d t
\end{multline}
Next, we make the change of variable
$$
r=\frac{2}{s\tau}\sqrt{\frac{ \varphi(t)}{t}}
$$
to obtain
\begin{multline}
M_{p,s}^{(m,n)}(\alpha) = c_{p,s}(\alpha)\left( \frac{2}{s} \right)^{m+n}
\int_{0}^\infty \int_0^\infty \frac{t^{p-m-1}}{\tau^{m+n+1}} \varphi^{n}(t) \left( \e^{\i \alpha} + t \e^{-\i \alpha} \right)^{m-n} \\
\times \exp\left \{- \frac{\tau}{2} - \frac{2\varphi(t)}{s^2\tau}\right \} \d \tau \d t  \\
=c_{p,s}(\alpha)\left( \frac{2}{s} \right)^{m+n}\int_{0}^\infty \int_0^\infty 
\frac{t^{p-m-1}}{\tau^{m+n+1}} \left [ t + 1/t + 2 \cos(2 \alpha) \right ]^{n} \left( \e^{\i \alpha} + t \e^{-\i \alpha} \right)^{m-n} \\
\times \exp\left\{-\frac{2}{s^2\tau} \left(t + \frac{1}{t} \right)\right \} \exp\left \{-\frac{\tau}{2} - \frac{4\cos(2\alpha)}{s^2\tau} \right \} \d t \d \tau\,.
\label{momentsFormula}
\end{multline}

By using the fact that
$$
K_q ( z ) = \frac{1}{2}\,\int_0^\infty t^{q-1} \exp \left \{ - \frac{z}{2} (t+1/t)
\right \}\,\d t,
$$
we can, provided that $n \le m$, express this double integral in terms of the integrals (\ref{macdonaldIntegral}).
For instance, when
$m = n = 0$,
Equation (\ref{momentsFormula}) yields
\begin{multline}
\notag
1 = M_{p,s}^{(0,0)}(\alpha)  \\ = c_{p,s} (\alpha) \int_0^\infty \exp \left \{ -\frac{\tau}{2} - \frac{4\cos(2\alpha)}{s^2\tau} \right \} \int_0^\infty t^{p-1} 
\exp\left\{-\frac{2}{s^2\tau} \left(t + \frac{1}{t} \right)\right \} \d t \frac{\d \tau}{\tau} \\
= 2 \,c_{p,s} (\alpha) \int_0^\infty \exp \left \{ -\frac{\tau}{2} - \frac{4\cos(2\alpha)}{s^2\tau} \right \} 
K_p \left ( \frac{4}{s^2 \tau} \right ) \frac{\d \tau}{\tau}\,.
\end{multline} 
This can be expressed  as
$$
1 = M_{p,s}^{(0)} (\alpha) = 4 \,c_{p,s}(\alpha) \,{\mathcal I}_p^{(-1)}\left ( u,v \right ), \quad
u = \overline{v} = 2 \e^{\i \alpha}/s\,.
$$
So we deduce the value of the normalisation constant from Equation (\ref{macdonaldFormula}).
In the same way, by taking $m=1$ and $n=0$, we obtain the mean:
\begin{equation}
\notag
{\mathbb E} \left ( Z \right ) = M_{p,s}^{(1,0)} (\alpha)
=  4\,c_{p,s}(\alpha) \left \{  v
{\mathcal I}_{p-1}^{(-2)}\left ( u, v \right )
+ u
{\mathcal I}_{p}^{(-2)}\left ( u,v \right )
\right \} \,.
\end{equation}
Its value is given by Equation (\ref{mean}). For the variance, we use
$$
\text{Var} \left ( Z \right ) = M_{p,s}^{(1,1)} (\alpha) - \left | M_{p,s}^{(1,0)}(\alpha)\right |^2\,.
$$
Equation (\ref{momentsFormula}) gives
\begin{equation}
\notag
M_{p,s}^{(1,1)} (\alpha)
= 4 \,c_{p,s}(\alpha) \left ( \frac{2}{s} \right )^2
\left \{ {\mathcal I}_{p-2}^{(-3)} (u,v) + 2 \cos(2 \alpha) \,{\mathcal I}_{p-1}^{(-3)} (u,v) +
{\mathcal I}_{p}^{(-3)} (u,v)
\right \}
\end{equation}
and so we deduce the value given in Equation (\ref{variance}).


\section{The Formula for the Lyapunov exponent}
\label{Lyap}


In this appendix, we consider matrices distributed according to a measure $\mu$ on $SL(2,\mathbb{C})$ of the form
\begin{equation*}
\label{matrix}
{\mathcal A} =\begin{bmatrix}
a & b \\
c & d \\
\end{bmatrix}
\end{equation*}
with exactly one row being random. In other words, the measure $\mu$ is such that one
or the other of the following conditions holds:
\begin{equation}
\tag{{\bf I}}
\text{$a$ and $b$ are fixed, $c$ and $d$ are random}
\label{caseI}
\end{equation}
or
\begin{equation}
\tag{{\bf II}}
\text{$c$ and $d$ are fixed, $a$ and $b$ are random}\,.
\label{caseII}
\end{equation}
 
Let $\nu$ be $\mu$-invariant measure on the projective space $\mathbb{P}(\mathbb{C}^2)=\bar{\mathbb{C}}$. 
In this case, we have (cf. Equation (1) in \cite{BoLa}, p. 9)
\begin{equation*}
\label{nu-inv}
\int_{\mathbb{C}} \int_{SL(2,\mathbb{C})}
\varphi ({\mathcal A}  \cdot z) d\mu(\mathcal A)d\nu(z) = \int_{\mathbb{P}(\mathbb{C}^2)}
\varphi(z) d\nu(z)
\end{equation*}
for every bounded Borel function $\varphi$.

First, let us consider the case where Condition (\ref{caseI}) holds. Assume
that $\nu$ has some positive moment. Suppose also that $\mu$ satisfies the
conditions required for the existence and positivity of the
Lyapunov exponent $\lambda$  (see Theorem 3.6 in \cite{BoLa}, p. 27---
which holds also in the complex domain). Then
\begin{multline}
\notag
\lambda=\int_{\mathbb{C}} \int_{SL(2,\mathbb{C})}
\ln \frac{ \left | {\mathcal A}  \binom{z}{1} \right |}{\left | \binom{z}{1} 
\right |} \,\d\mu(\mathcal A) \,\d\nu(z)=\\
\frac{1}{2} \int_{\mathbb{C}} \int_{SL(2,\mathbb{C})}
\ln \frac{ |az+b|^2 +|cz+d|^2}{ |z|^2 +1} \,\d \mu(\mathcal A) \,\d \nu(z)=\\
\frac{1}{2} \int_{\mathbb{C}} \int_{SL(2,\mathbb{C})}
\ln\left(1+|\mathcal{A}\cdot z|^{-2} \right)+
\ln \frac{ |az+b|^2 }{|z|^2 +1} \,\d \mu(\mathcal A) \,\d \nu(z)=\\
\frac{1}{2} \int_{\mathbb{C}}
\ln\left(1+| z|^{-2} \right) \,\d \nu(z)
+\int_{\mathbb{P}(\mathbb{C}^2)} \ln\left|a+bz^{-1} \right | \,\d \nu(z)\\-
\frac{1}{2} \int_{\mathbb{C}}
\ln\left(1+| z|^{-2} \right ) \,\d \nu(z)=
\int_{\mathbb{C}} \ln\left|a+bz^{-1} \right|\, \d\nu(z).
\end{multline}
We note that the splitting of the integrals in this calculation is permissible because 
the assumed existence of a positive
moment of $\mu$ guarantees the integrability of $\ln(1+|z|^{-2})$ and $\ln(a+b|z|^{-1})$.

The case where Condition (\ref{caseII}) holds involves a similar calculation; hence
\begin{equation}
\label{GLE}
{\lambda}=
\begin{cases}
\int_{\mathbb{C}} \ln|a+bz^{-1}| \,\d {\nu}( z)  & \text{if Condition  (\ref{caseI}) holds} \\
& \\
\int_{\mathbb{C}} \ln|cz+d|  \,\d {\nu}( z) & \text{if Condition  (\ref{caseII}) holds}
\end{cases}\,.
\end{equation}

Now, we apply this result to the matrices (\ref{randomMatrix}) considered in Section \ref{prelim}.
In this case, Condition (\ref{caseI}) holds; there is only one random entry, and
since it is gamma-distributed, the support of $\mu$ is not contained in any
compact subgroup of $\text{GL}(2,{\mathbb C})$. 
Furthermore 
$$
{\mathcal A}^{*}_{n}(\alpha)={\mathcal A}_{n}(-\alpha)
$$ 
and hence the invariant measure for ${\mathcal A}^{*}_{n}(\alpha)$ is continuous for all $\alpha \in [-\pi/2, \pi/2]$.
We may therefore use Proposition 3.3 in \cite{BoLa}, p. 26, to deduce the uniqueness of the invariant measure $\nu$ found in the paper. Since, for all $\alpha\in [-\pi/2, \pi/2]$, this measure possesses positive moments of all orders less then one, 
we can apply Formula (\ref{GLE}) to obtain (\ref{lyapunovIntegral}).

For matrices of the form (\ref{umatrix}) and for Schr\"odinger matrices, it is
Condition (\ref{caseII}) that holds, and the formula (\ref{GLE}) 
leads to (\ref{uLE}).

\section{Proof of Theorem \ref{generalisedLetacSeshadriThm}}
\label{generalisedLetacSeshadriAppendix}

\begin{proof}
By direct substitution in Equation (\ref{equation4u}).
To facilitate the calculation, define
\begin{multline}
\notag
r_a = \sqrt{(x-a\cos \alpha)^2 + (y-a\sin\alpha)^2},
\\
X_a =
-\frac{(x-a \cos \alpha)
\sin \alpha+(y-a \sin \alpha) \cos \alpha}{r_a^2}
\quad \text{and} \\
Y_a = (y-a \sin \alpha) \cos \alpha -
(x-a \cos \alpha) \sin \alpha \,.
\end{multline}
We note that
$$
r_0 = r, \quad X_0 = X \quad \text{and} \quad Y_a \equiv Y\,.
$$

We can write
\begin{multline}
f_\alpha(z-a \e^{\i \alpha}) \gamma_{p,s}(a)
= C Y^{p-1} \exp \left \{ \frac{\sin(2 \alpha)}{s}
\frac{1}{Y} \right \} \\
\times \frac{a^{p-1}}{s^p \Gamma(p)}
\left [ r_a^2 X_a \right ]^{-(p+1)} \exp \left \{ E(a)/s \right \},
\label{muTimesnu}
\end{multline}
where
$$
E(a) = \frac{\sin(2 \alpha)}{X_a} - a\,.
$$
A lengthy but straightforward calculation reveals that
$$
E'(a) = -\left [ \frac{Y}{r_a^2 X_a} \right ]^2\,.
$$
Hence, by integrating Equation (\ref{muTimesnu}) with respect to
$a$, we obtain
\begin{equation}
\int_0^{a(x,y)}
f_\alpha(z-a \e^{\i \alpha}) \gamma_{p,s}(a) \,\d a
= C \frac{Y^{p-1}}{Y^2} \exp \left \{ \frac{\sin(2 \alpha)}{s}
\frac{1}{Y} \right \}  \,{\mathcal E}_p,
\label{leftHandSide}
\end{equation}
where
$$
{\mathcal E}_p = -\int_0^{a(x,y)}
\frac{1}{s^p \Gamma(p)} \left [ \frac{a}{r_a^2 X_a}
\right ]^{p-1} E'(a)
\exp \left \{ E(a)/s \right \}\,\d a\,.
$$
Noting that
\begin{equation}
\notag
\frac{\d}{\d a} \left [ \frac{a}{r_a^2 X_a}
\right ]
= -\frac{x \sin \alpha+y \cos \alpha}{r_a^4 X_a^2}
= \frac{r^2 X}{r_a^4 X_a^2}
= -\frac{r^2 X}{Y^2} E'(a),
\end{equation}
we obtain, after integration by parts,
\begin{multline}
\notag
{\mathcal E}_p =  \left .
\frac{-1}{s^{p-1} \Gamma(p)} \left [ \frac{a}{r_a^2 X_a}
\right ]^{p-1}
\exp \left \{ E(a)/s \right \} \right |_{0}^{a(x,y)}
+
\begin{cases}
0 & \text{if $p=1$} \\
\frac{r^2 X}{Y^2} {\mathcal E}_{p-1} & \text{if $p>1$}
\end{cases} \\
= \begin{cases}
\exp \left \{ E(0)/s \right \} & \text{if $p=1$} \\
\frac{r^2 X}{Y^2} {\mathcal E}_{p-1} & \text{if $p>1$}
\end{cases}
\,.
\end{multline}
Thus,
$$
{\mathcal E}_p = \left [ \frac{r^2 X}{Y^2} \right ]^{p-1} \,\exp \left \{
\frac{\sin(2 \alpha)}{s} \frac{1}{X}
\right \}
$$
and, reporting this in Equation (\ref{leftHandSide}),
we have shown that
\begin{multline}
\notag
\int_0^{a(x,y)}
f_\alpha(z-a \e^{\i \alpha})
\gamma_{p,s}(a) \,\d a  \\
=  \frac{C}{Y^2} \left ( \frac{r^2 X}{Y}\right )^{p-1}
\exp \left \{ \frac{\sin(2 \alpha)}{s}
\left [ \frac{1}{Y} + \frac{1}{X} \right ] \right \}\,.
\end{multline}
It follows easily that $f_\alpha$ satisfies Equation
(\ref{equation4u}). 
\end{proof}

\section{Proof of Theorem \ref{alphaIsPiOver2Theorem}}
\label{P2Thm}
\begin{proof}
We need only consider the case $\alpha = \pi/2$.
As in the case $|\alpha| < \pi/2$, the proof is by direct substitution
of the expression
\begin{equation}
\notag
f_{+}(y) = 
\frac{1}{y^{p+1}} 
\exp \left [  \frac{1}{s} 
\left ( \frac{1}{y} -y \right ) \right ] 
\int_{c(y)}^y 
\exp \left [ -\frac{1}{s} 
\left ( \frac{1}{t} -t\right ) \right ]\,t^{p-1}\,\d t\,,
\end{equation} 
where
$$
c(y) = \begin{cases}
-\infty & \text{if $y<0$} \\
0 & \text{if $y>0$}
\end{cases},
$$
into the integral equation (\ref{equation4fWhenAlphaIsPiOver2}). 

By making the substitution $u = -1/y-a$ in
the integral on
the right-hand side of Equation (\ref{equation4fWhenAlphaIsPiOver2}), we
obtain
\begin{multline}
\frac{1}{y^2} \int_0^\infty f_{+} \left ( -\frac{1}{y}-a 
\right ) \gamma_{p,s}(a) \d a \\
= \frac{1}{y^2} \int_{-\infty}^{-1/y}
f_{+} (u) \gamma_{p,s} \left ( -\frac{1}{y}-u \right ) \d u \\
= \frac{1}{y^2} \int_{-\infty}^{-1/y} \int_{c(u)}^u \frac{1}{u^{p+1}}
\exp \left [ \frac{1}{s} \left ( \frac{1}{u} - \frac{1}{t}+t
+ \frac{1}{y} \right ) \right ] \\
\times t^{p-1}
\frac{1}{s^p \Gamma(p)} \left (-\frac{1}{y} -u \right )^{p-1}
 \d t \d u\,. 
\label{rightHandSide}
\end{multline}
At this point, it is helpful to treat the cases $y<0$ and $y > 0$
separately. 

Let us begin by assuming that $y>0$. Then $c(u) = -\infty$
on the right-hand side of
the last equation and, by changing the order of integration,
we obtain
\begin{multline}
\notag
\frac{1}{y^2} \int_0^\infty f_{+} \left ( -\frac{1}{y}-a 
\right ) \gamma_{p,s}(a) \d a \\
= \frac{\e^{\frac{1}{sy}}}{y^{p+1}}  \int_{-\infty}^{-1/y}
\exp \left [
-\frac{1}{s} \left ( \frac{1}{t}-t \right ) \right ] \\
\times t^{p-1}
\int_{t}^{-1/y} \frac{1}{s^p \Gamma(p)} \left ( - \frac{1}{u}-y
\right )^{p-1} \e ^{\frac{1}{s u}} \frac{1}{u^2} \d u \d t\,.
\end{multline}
Then, after making the substitution $\tau = -1/t$,
followed by $v= 1/u+y$, this becomes
\begin{multline}
\notag
\frac{1}{y^2} \int_{0}^\infty f_{+} \left ( -\frac{1}{y}-a \right )
\gamma_{p,s}(a) \d a \\
= \frac{1}{y^{p+1}} \exp \left [ \frac{1}{s} \left ( \frac{1}{y}-y
\right ) \right ] \int_0^y \exp \left [ -\frac{1}{s} \left (\frac{1}{\tau}
- \tau \right ) \right ] \frac{1}{\tau^{p+1}} \varphi (y-\tau) \d \tau, 
\end{multline}
where
\begin{equation}
\varphi (w) = \int_0^{w} \frac{1}{s^p \Gamma (p)} v^{p-1} 
\e^\frac{v}{s} \d v\,.
\label{phi}
\end{equation}
In order to prove that Equation (\ref{equation4fWhenAlphaIsPiOver2})
holds for $y>0$, we therefore only need to verify that
\begin{equation}
\notag
\int_0^y \exp \left [ -\frac{1}{s} \left ( \frac{1}{t} - t \right ) \right ]
t^{p-1} \d t 
= 
\int_0^y \exp \left [ -\frac{1}{s} \left ( \frac{1}{t} - t \right ) \right ]
\frac{1}{t^{p+1}} \varphi (y-t)  \d t
\end{equation}
holds identically for every $y>0$. Clearly, it holds
when $y = 0+$, and so it will be sufficient to show that the {\em derivatives}
are the same. The derivative of the right-hand side
is
\begin{multline}
\notag
\exp \left [ -\frac{1}{s} \left ( \frac{1}{y}-y \right ) \right ]
\frac{1}{y^{p+1}} \varphi(0)
+ \int_0^y \exp \left [ -\frac{1}{s} \left ( \frac{1}{t} - t 
\right ) \right ] \frac{1}{t^{p+1}} \varphi'(y-t) \d t \\
= \int_0^y \exp \left [ -\frac{1}{s} \left ( \frac{1}{t} - t 
\right ) \right ] \frac{1}{t^{p+1}} \frac{1}{s^p \Gamma(p)}
(y-t)^{p-1} \exp \left [ \frac{1}{s} (y-t) \right ] \d t \\
= y^{p-1} \e^{\frac{y}{s}} \int_0^y \frac{1}{s^p \Gamma (p)}
\left ( \frac{1}{t} - \frac{1}{y} \right )^{p-1} \e^{-\frac{1}{s t}}
\frac{1}{t^2} \d t \\
= y^{p-1} \exp \left [ - \frac{1}{s} \left ( \frac{1}{y}
- y \right ) \right ] \int_0^\infty \frac{1}{s^p \Gamma(p)}
u^{p-1} \e^{-\frac{u}{s}} \d u,
\end{multline}
after making the substitution $u = 1/t-1/y$. The result follows.

To complete the proof, we also need to consider the case $y<0$. Returning
to Equation (\ref{rightHandSide}), we need to split the range of
integration of the variable $u$ into positive and negative values.
Recalling the definition of $c(u)$ and, as before, changing the order
of integration, we obtain
\begin{multline}
\notag
\frac{1}{y^2} \int_0^\infty f_{+} \left ( -\frac{1}{y}-a 
\right ) \gamma_{p,s}(a) \d a \\
= \frac{\e^{\frac{1}{sy}}}{y^{p+1}}  \int_{-\infty}^{0}
\exp \left [
-\frac{1}{s} \left ( \frac{1}{t}-t \right ) \right ] \\
\times t^{p-1}
\int_{t}^{0} \frac{1}{s^p \Gamma(p)} \left ( - \frac{1}{u}-y
\right )^{p-1} \e ^{\frac{1}{s u}} \frac{1}{u^2} \d u \d t \\
= \frac{\e^{\frac{1}{sy}}}{y^{p+1}}  \int_{0}^{-1/y}
\exp \left [
-\frac{1}{s} \left ( \frac{1}{t}-t \right ) \right ] \\
\times t^{p-1}
\int_{t}^{-1/y} \frac{1}{s^p \Gamma(p)} \left ( - \frac{1}{u}-y
\right )^{p-1} \e ^{\frac{1}{s u}} \frac{1}{u^2} \d u \d t
\,.
\end{multline}
After making the substitution $v = -1/u -y$, this becomes
\begin{multline}
\notag
\frac{1}{y^2} \int_0^\infty f_{+} \left ( -\frac{1}{y}-a 
\right ) \gamma_{p,s}(a) \d a \\
= \frac{1}{y^{p+1}} \exp \left [ \frac{1}{s} \left ( \frac{1}{y}
- y \right ) \right ]  
\int_{-\infty}^{-1/y} \exp \left [ -\frac{1}{s}
\left ( \frac{1}{t}-t \right ) \right ] \phi(-1/t-y)
t^{p-1} \d t \\
- 
\frac{1}{y^{p+1}} \exp \left [ \frac{1}{s} \left ( \frac{1}{y}
- y \right ) \right ]  
 \int_{0}^{-1/y} \exp \left [ 
-\frac{1}{s} \left ( \frac{1}{t}-t \right ) \right ] t^{p-1} \d t,
\end{multline}
where
$$
\phi(w) = \int_w^\infty \frac{1}{s^p \Gamma(p)} v^{p-1} \e^{-\frac{v}{s}}
\d v\,.
$$
Therefore, we have to prove that 
\begin{multline}
\notag
\int_{-\infty}^y \exp \left [-\frac{1}{s} 
\left ( \frac{1}{t}-t\right )  \right ] t^{p-1} \d t \\ =
\int_{-\infty}^{-1/y} \exp \left [ -\frac{1}{s}
\left ( \frac{1}{t}-t \right ) \right ] \phi(-1/t-y)
t^{p-1} \d t \\
- \int_{0}^{-1/y} \exp \left [ 
-\frac{1}{s} \left ( \frac{1}{t}-t \right ) \right ] t^{p-1} \d t
\end{multline}
holds for every $y<0$ . On both sides,
the limit as $y \rightarrow -\infty$ is zero. 
Furthermore, the derivative
of the right-hand side is
\begin{multline}
\notag
\exp \left [ - \frac{1}{s} \left ( \frac{1}{y}-y 
\right ) \right ] \frac{(-1)^{p-1}}{y^{p+1}} \left [ \phi (0)-1 \right ] \\
- \int_{-\infty}^{-1/y} \exp \left [ -\frac{1}{s} \left ( 
\frac{1}{t}-t \right )
\right ] t^{p-1} \phi'(-1/t-y) \\
= \int_{-\infty}^{-1/y} \exp \left [ -\frac{1}{s} \left ( \frac{1}{t}
-t \right ) \right ] t^{p-1} \frac{1}{s^p \Gamma(p)} 
\left ( -\frac{1}{t}-y \right )^{p-1} \exp \left [ -\frac{1}{s}
\left ( -\frac{1}{t}-y \right ) \right ] \d t \\
= y^{p-1} \e^{\frac{y}{s}} \int_{-\infty}^{-1/y} \frac{1}{s^p \Gamma(p)}
\left ( -\frac{1}{y}-t\right )^{p-1} \e^{\frac{t}{s}} \d t \\
= y^{p-1} \exp \left [ -\frac{1}{s} \left ( \frac{1}{y}-y
\right ) \right ] \int_0^\infty \frac{1}{s^p \Gamma(p)}
v^{p-1} \e^{-v} \d v
\end{multline}
after making the substitution $v = -1/y-t$. 
\end{proof}

\section{An asymptotic expansion for $f_{+}$ as $s \rightarrow 0$}
\label{lyapunovAppendix}

It is actually simpler to work with 
$$
u(y) =  \frac{1}{y^2} f_{+}(-1/y)\,.
$$
So let write
$$
u(y) \sim 
\sum_{n=1}^\infty u^{(n)}(y) s^n \quad \text{as $s \rightarrow 0$}\,.
$$
where
$$
u^{(n)}(y) = \frac{1}{y^2} f_{+}^{(n)}(-1/y)\,. 
$$
Equation (\ref{recurrence4fn}) then gives
\begin{equation}
u^{(1)} (y) = \frac{1}{1+y^2} \quad \text{and} \quad 
u^{(n+1)}(y) = \frac{-y^{p+1}}{1+y^2} \frac{\d }{\d y} \left [ \frac{u^{(n)}(y)}{y^{p-1}} \right ]
\,.
\label{recurrence4un}
\end{equation}
A direct calculation of the first few terms reveals that the
$u^{(n)}$ have a very simple structure:
\begin{equation}
u^{(2n-1)}(y) = \sum_{k=n}^{4n-3} \frac{u_k^{(2n-1)}}{(1+y^2)^k}
\quad \text{and} \quad 
u^{(2n)}(y) = \sum_{k=n+1}^{4n-1} \frac{u_k^{(2n)}y}{(1+y^2)^k}, \quad u_k^{(n)}
\in {\mathbb R}\,.
\label{theUn}
\end{equation}
By substituting these expressions in
Equation (\ref{recurrence4un}), one obtains the recurrence relations
\begin{align*}
u_{n+1}^{(2n)} &= (p-1+2n) u_{n}^{(2n-1)}, \\
u_{k}^{(2n)} &= (p-3+2k) u_{k-1}^{(2n-1)} - 2(k-2) u_{k-2}^{(2n-1)}, 
\quad n+2 \le k \le 4n-2, \\
u_{4n-1}^{(2n)} &= -2 (4n-3) u_{4n-3}^{(2n-1)}, \\
u_{n+1}^{(2n+1)} &= (p+2n) u_{n+1}^{(2n)}, \\
u_{n+2}^{(2n+1)} &= (p+2n+2) u_{n+2}^{(2n)} - (p+2+4n) u_{n+1}^{(2n)}, \\
u_k^{(2n+1)} &= (p-2+2k)u_k^{(2n)}  \\
             &+ (6-p-4k) u_{k-1}^{(2n)}+ 2(k-2) u_{k-2}^{(2n)},\quad n+3 \le k \le 4n-1, \\
u_{4n}^{(2n+1)} &= (6-p-16n) u_{4n-1}^{(2n)} + 2(4n-2) u_{4n-2}^{(2n)}, \\
u_{4n+1}^{(2n+1)} &= (8n-2) u_{4n-1}^{(2n)}, 
\end{align*}
with the starting value $u_1^{(1)} = 1$.
We then have
\begin{equation}
\int_{{\mathbb R}} f_{+}(y) \d y = \int_{{\mathbb R}} u(y) \d y 
\sim 2 \sum_{n=1}^\infty s^{2n-1} \sum_{k=n}^{4n-3} u_{k}^{(2n-1)} \alpha_k
\label{denominatorSeries}
\end{equation}
and
\begin{equation}
-\int_{\mathbb R} \ln |y| \,f_{\pi/2}(y) \d y = \int_{\mathbb R} \ln |y| \,u(y) \d y 
\sim 2 \sum_{n=1}^\infty s^{2n-1} \sum_{k=n}^{4n-3}
u_{k}^{(2n-1)} \beta_k
\label{numeratorSeries}
\end{equation}
as $s \rightarrow 0$,
where 
$$
\alpha_{k} = \int_0^\infty \frac{\d y}{(1+y^2)^k} \quad \text{and} \quad
\beta_k = \int_0^\infty \frac{\ln y\, \d y}{(1+y^2)^k}\,.
$$
The $\alpha_k$ and $\beta_k$ satisfy the recurrence relations
$$
\alpha_{k+1} = \left ( 1- \frac{1}{2k} \right ) \alpha_k, \quad 
\beta_{k+1} = \beta_k - \frac{\alpha_k+\beta_k}{2 k}, \quad 
\alpha_1 = \frac{\pi}{2}, \;\beta_1 = 0\,.
$$

\end{document}